\documentclass[11pt,a4paper]{article}
\pdfoutput=1

\usepackage{jheppub}
\usepackage{mathtools}
\usepackage{dsfont}
\usepackage{braket}
\newcommand\numberthis{\addtocounter{equation}{1}\tag{\theequation}}

\newcommand{\DD}{\mathrm{d}}
\renewcommand{\.}{\cdot}
\newcommand{\defined}{\coloneqq}
\newcommand{\e}{\epsilon}
\renewcommand{\d}{\partial}
\renewcommand{\t}[1]{\text{#1}}
\newcommand{\at}[1]{\bigg\vert_{#1}}

\renewcommand{\O}{\mathcal{O}}

\title{A perturbation theory for the Coulomb-phase infrared divergence}

\author[a]{Luke Lippstreu}

\affiliation[a]{Higgs Centre, School of Physics \& Astronomy \\
		University of Edinburgh, EH9 3FD, United Kingdom}
\emailAdd{llippstr@ed.ac.uk}

\abstract{We construct a perturbation theory which we conjecture to be free of the Coulomb-phase infrared divergence. This perturbation theory is developed for one of the simplest yet prototypical scattering amplitudes which would otherwise exhibit this divergence: the semiclassical scattering of a spinless boson on a background Coulomb field. The perturbation theory is based upon replacing plane waves with Coulomb wavefunctions, and the free-field propagator with the Coulomb propagator, in order to appropriately match the asymptotics of the exact in/out states. We compute the leading-order (LO) and next-to-leading-order (NLO) infrared-finite scattering amplitudes in this framework, which include effects to all orders in the coupling, and demonstrate that these amplitudes are in agreement with the known exact amplitude at these orders. We comment on the Runge-Lenz symmetry of the LO amplitude, under which the states furnish a principal-series representation of the Euclidean conformal group on the 2-sphere.}

\begin{document}

\maketitle
\section{Introduction}
Infrared divergences occur in interacting theories that contain massless particles in four or fewer spacetime dimensions, and are thus of considerable importance for the Standard Model of particle physics. These divergences can be alleviated at the level of inclusive cross sections for sufficiently simple theories like QED \cite{Bloch:1937pw}. However, in order to study questions regarding causality and unitarity, it is preferable to alleviate these divergences at the level of amplitudes, ideally without introducing an IR cutoff. The Faddeev-Kulish (FK) framework \cite{Kulish:1970ut,Dollard1964AsymptoticCA,Kibble:1968sfb,Chung:1965zza} represents one solution to the problem of constructing an IR-finite S-matrix, and while it has been useful for uncovering many features of the infrared structure of gauge theories and gravity, it has rarely been used for the explicit calculation of IR-finite scattering amplitudes, with \cite{Hannesdottir:2019opa,Hannesdottir:2019umk,Forde:2003jt} being the only instances of explicitly calculated amplitudes we were able to find.\par
One of the simplest examples of an infrared divergence is encountered when using standard techniques to compute the semiclassical scattering of a Klein-Gordon (KG) particle off a Coulomb background \cite{kang1962higher}. In this paper, we construct a perturbation theory, conjectured to be IR-finite, for this semiclassical amplitude and calculate it up to NLO. The LO (\ref{eq:LOcovamp}) and NLO (\ref{NLO spherical}) amplitudes we compute are IR-finite and do not contain any implicit or explicit cutoff or dimensional regularization scale. The exact IR-finite amplitude (\ref{eq:exactamp}) is known by other non-perturbative means, and we find agreement with this result up to the orders we compute, thus providing confidence for extending the perturbation theory to less trivial settings where the exact answer is not known. \par
The underlying cause of infrared divergences, both in the semiclassical and the quantum cases, is that states in such theories do not appropriately converge towards non-interacting free particle states at asymptotic times, which is counter to an assumption of standard perturbation theory. This does not imply that scattering amplitudes are ambiguous, zero, infinite, or cutoff dependent; it only implies that one should not use a perturbation theory that assumes asymptotically non-interacting states. In the semiclassical case, it is this assumption regarding asymptotically non-interacting particles which leads to the use of plane waves as a useful basis of wavefunctions for perturbatively constructing the amplitude. In the presence of long-range forces it would be preferable to use a basis of wavefunctions whose asymptotics appropriately match those of the exact in/out states. In this paper we develop a perturbation theory which uses a basis of Coulomb wavefunctions whose asymptotics appropriately match those of the exact in/out states; namely, the perturbation theory implements the replacement
\begin{gather}
    e^{-ip\.x}\rightarrow e^{-ip\.x}{}_1F_1\Big(-i\frac{e_1e_2}{4\pi}\frac{E}{|\vec{p}|},1,i(|\vec{p}||\vec{x}|-\vec{p}\.\vec{x})\Big)\label{eq:main idea}
\end{gather}
(the Lorentz-invariant form of the RHS is given in (\ref{eq:instate}, \ref{eq:outstate})) where ${}_1F_1(a,b,z)$ is the confluent hypergeometric function, $e_1e_2$ are the charges of the particles, and when we turn off the coupling we recover standard plane waves, since ${}_1F_1(0,b,z)=1$. Let us be quick to emphasize that when implementing the replacement (\ref{eq:main idea}) we will not be redefining the scattering amplitude, nor will we be scattering wave packets. The infrared-finite semiclassical scattering amplitude is unambiguously defined, as we review in section \ref{sect:setup}, without any reference to the perturbation theory one uses to compute it. Whether one uses plane waves or Coulomb wavefunctions is a matter of framework. If one uses plane waves, then in order to account for the disagreement between the asymptotics, one has to implement the Faddeev-Kulish approach by modifying the evolution operators. Instead, if we use Coulomb wavefunctions, then the asymptotics agree, and the modification to the evolution operators is much simpler. Both approaches should give the same answer, as they are both computing a quantity that is independent of the perturbative framework. To demonstrate this equivalence, in section \ref{sect:summary} we argue that the relevant Faddeev-Kulish Moller operators act as projection operators which implement the mapping (\ref{eq:main idea}). \par
Infrared divergences make the causality and unitarity properties of observables in QFT less transparent. For example, many of the most useful theorems concerning the analytic structure of S-matrices in $4D$, such as the Froissart-Martin bound and the Steinmann relations \cite{Steinmann1960}, require the existence of a mass gap, which is somewhat disappointing given that QED and gravity do not have a mass gap. Without firm control over the cutoff-independent IR-finite S-matrix, it is significantly more difficult to study the analytic structure of observables in gapless $4D$ QFTs and deduce the analogs of these theorems \cite{Bourjaily:2020wvq, Fevola:2023fzn,Berghoff:2022mqu,Hannesdottir:2022xki}. Properties such as these are of central importance in the modern-day S-matrix bootstrap program \cite{deRham:2022hpx,Kruczenski:2022lot}, as well as for a variety of modern-day techniques for calculating observables in QFT \cite{caronhuot2020steinmann}; thus, it would be beneficial to have a framework which gives IR-finite, cutoff-independent, unambiguous scattering amplitudes. \par
There are three distinct phenomena which are generally referred to as infrared divergences: collinear, real-emission, and imaginary (or Coulomb-phase) divergences. This article only addresses the last of these. Collinear divergences arise in theories which contain vertices where three massless particles can meet. We will not study such theories in this paper. The distinction between real and imaginary infrared divergences can be explained by looking at the soft exponentiation factor for the leading infrared divergences in abelian gauge theories \cite{yennie,Weinberg:1965nx}. This theorem states that the leading IR divergences in an abelian gauge theory can be factored out of the scattering amplitude $\mathcal{A}$ (see Eq. 13.2.4 of \cite{Weinberg:1995mt} for a textbook discussion)
\begin{gather}
    \mathcal{A}=\Bigg(\frac{\Lambda_{\text{IR}}}{\lambda_{\text{IR}}}\Bigg)^{R+I}\mathcal{A}_0(\Lambda_{\text{IR}}),\label{eq:Weinbergexponentiation}
    \end{gather}
    where the real and imaginary parts are respectively 
    \begin{gather}
    R=\frac{1}{16\pi^2}\sum_{i,j}\frac{\eta_i\eta_je_ie_j}{\beta_{ij}}\log\Big(\frac{1+\beta_{ij}}{1-\beta_{ij}}\Big)\quad,\quad I=\frac{1}{8\pi}\sum_{\eta_i\eta_j=+1}-i\frac{e_ie_j}{\beta_{ij}} \numberthis\label{eq:Weinbergexponential},\\
    \beta_{ij}\defined \sqrt{1-\frac{m_i^2m_j^2}{(p_i\.p_j)^2}},
\end{gather}
where $\eta=+1/-1$ corresponds to in/out states, respectively, $\lambda_{\text{IR}}/\Lambda_{\text{IR}}$ represent the lower/upper infrared (IR) cutoff scales for the energies of exchanged photons, and $\mathcal{A}_0(\Lambda_{\text{IR}})$ only contains photon exchanges with energies greater than $\Lambda_{\text{IR}}$. The real factor $R$ in (\ref{eq:Weinbergexponential}) is related to the production of real photons, as it is this term which cancels out at the level of inclusive cross sections when summing over soft radiative processes. We will not consider this term any further, and explain in Sect. \ref{sect:setup} how, by using a retarded propagator for the gauge fields, we are able to avoid dealing with this term, so that we can isolate the focus of this work: the Coulomb-phase divergence. The second term in (\ref{eq:Weinbergexponential}) is referred to as the Coulomb-phase divergence, and it is purely imaginary. The $\eta_i\eta_j=+1$ indicates that this term only occurs for pairs of particles that are both in-states or both out-states. Physically, one can associate the Coulomb-phase divergence with the fact that the late-time trajectories of classical particles interacting via a $\frac{1}{r}$ potential do not asymptote to free-particle trajectories, but rather to hyperbolic trajectories which have a logarithmic correction to free-particle motion (see, e.g., \cite{PhysRevD.7.1082,Sahoo_2019} for the trajectory, and appendix H of \cite{Hirai:2021ddd} for the relation between the trajectory and this phase). \par
There have been several studies on alleviating infrared divergences at the level of scattering amplitudes. In \cite{Duary:2022afn,Duary:2022pyv,Hijano:2020szl} the possibility of obtaining an IR-finite S-matrix from the flat-space limit of AdS was explored. In \cite{Feal:2022iyn,Feal:2022ufw,Bonocore:2020xuj} a many-body theory of worldlines was developed that allows for the easy identification and removal of IR divergences. \cite{Prabhu:2022zcr} offers an up-to-date review of the relation between Faddeev-Kulish states and memory effects, why the framework fails for gauge theories with massless matter, as well as an outline for an algebraic framework for IR-finite scattering. An up-to-date review of techniques for calculating infrared-finite observables in gauge theory can be found in \cite{Agarwal:2021ais}. In \cite{Rai:2021vvq} a dynamical resummation method was developed to study the time evolution of infrared dressing. \cite{Hirai:2020kzx} argues that the FK states are not enough to cure IR divergences, and provides dressings which resolve this issue, with \cite{Furugori:2020vdl} providing yet another resolution to this problem by introducing a large-timescale cutoff. The FK formalism was extended to quantum gravity in \cite{Ware_2013}. The Faddeev-Kulish dressings have been useful for uncovering many aspects of the so-called infrared triangle of soft theorems, asymptotic symmetries, and memory effects \cite{Hirai:2022yqw, Gonzo:2022tjm, Pasterski:2022djr, Nguyen:2023ibj, Choi:2017ylo, Freidel:2022skz, Krishna:2023fxg, Kapec:2021eug, Kalyanapuram:2021tnl, Nguyen:2021ydb, Magnea:2021fvy, Arkani-Hamed:2020gyp, He:2020ifr, AtulBhatkar:2019vcb, Kapec:2017tkm, Ashtekar:2014zsa, Gabai:2016kuf, Choi:2017bna}, as well as entanglement properties of soft massless particles \cite{Su:2023ciz, Su:2023teb,Irakleous:2021ggq}, black-hole information \cite{Mirbabayi:2016axw, Bousso:2017dny, Strominger:2017aeh}, and many other important features of the infrared structure of gauge theories and gravity \cite{Csaki:2022tvb,Mund:2021zhx,Donnelly:2015hta,DiVecchia:2022piu,Nabeebaccus:2022jhu,Biswas:2022lsj,Bhamre:2022kmd,Cristofoli:2021jas,Berezhiani:2021zst, Boyanovsky:2021ija,Addazi:2021bqu, DeLisle:2022pjo,Gass:2021idx,Kolekar:2021bfa, Gonzo:2020xza,Ciafaloni:2018uwe, Carney:2017jut,Addazi:2019mjh, Frye:2018xjj}. As the Coulomb wavefunction perturbation theory studied in this paper offers an alternative to the Faddeev-Kulish framework, it would be interesting to study these features of gauge theories and gravity from this complementary perspective.
\par
We highlight that the perturbation theory developed in this paper can be considered a manifestly Lorentz-invariant generalization of the distorted-wave Born approximation \cite{Bethe,CROTHERS1992287,Glauber}.
\par
\paragraph{Outline} The paper is organized as follows. In section \ref{sect:reviewNRQM} we review the exact solution to the NRQM Coulomb scattering problem to explain the relevance of the Coulomb wavefunctions used throughout this paper, and to review why IR divergences artificially arise in standard perturbation theory. In section \ref{sect:reviewrunge} we review the Runge-Lenz symmetry of the NRQM amplitude to provide background for the Runge-Lenz symmetry observed in the LO amplitude described in section \ref{sect:runge lenz}. In section \ref{sect:setup} we provide the definition of the scattering amplitude which we compute in the rest of the paper. There, the semiclassical amplitude is defined to be the inner product of the in and out solutions to the KG equation in a background Coulomb field. In section \ref{sect:IR finite perturbation theory} we construct a conjecturally IR-finite perturbation theory for computing this amplitude. In section \ref{sect.Amplitudes} we compute the IR-finite LO (\ref{eq:LOcovamp}) and NLO (\ref{NLO spherical}) amplitudes in this framework. In section \ref{sect:exactresult} we compare these amplitudes to the known exact amplitude (\ref{eq:exactamp}) and find that they are in agreement. Section \ref{sect:conc} discusses future directions for adapting the perturbation theory to less trivial settings. Appendix \ref{app:exactamp} computes the exact IR-finite amplitude by non-perturbative means and also verifies that the asymptotics of the Coulomb wavefunctions appropriately match the asymptotics of the exact in/out states. Appendix \ref{app:LO Int} computes integrals and summations relevant for the LO and NLO amplitudes. Appendix \ref{app:Greensfunction} proves properties of the relativistic Coulomb Green's function used in the text.
\section{Review}
\subsection{NRQM Coulomb scattering}\label{sect:reviewNRQM}
The non-relativistic quantum mechanical (NRQM) scattering of a particle on a Coulomb potential is an exactly solvable problem that provides useful insights regarding infrared divergences. This section provides some of the properties of the Coulomb wavefunctions that we will use throughout this paper, and also motivates our search for a relativistic perturbation theory that has no IR divergences or cutoffs at any order of the coupling expansion.\par
The Coulomb Hamiltonian
\begin{gather}
    H=\frac{p^2}{2m}+\frac{e_1e_2}{4\pi}\frac{1}{r}\label{eq:CoulombHamiltonian}
\end{gather}
can be exactly solved. The exact in/out states with eigenvalue $E=\frac{|\vec{p}|^2}{2m}$ are \cite{Landau:1991wop}
\begin{align}
\phi_{\text{in}}(\vec{x},\vec{p})&=\Gamma(1+i\gamma)e^{-\frac{\pi\gamma}{2}}e^{-iEt+i\vec{p}\.\vec{x}}{}_1F_1\Big(-i\gamma,1,i(|\vec{p}||\vec{x}|-\vec{p}\.\vec{x})\Big)\label{NRQMinstate}\\
    \phi_{\text{out}}(\vec{x},\vec{p})&=\Gamma(1-i\gamma)e^{-\frac{\pi\gamma}{2}}e^{-iEt+i\vec{p}\.\vec{x}}{}_1F_1\Big(i\gamma,1,-i|\vec{p}||\vec{x}|-i\vec{p}\.\vec{x}\Big)\label{NRQMoutstate}
    \end{align}
    \begin{equation}
        \gamma_{\text{}}(p)\defined \frac{e_1e_2 m}{4\pi |\vec{p}|},\qquad E=\frac{|\vec{p}|^2}{2m}
    \end{equation}
 where ${}_1F_1(a,b,z)$ denotes the confluent hypergeometric function. Using the large $z$ asymptotics of this function
    \begin{gather}
       \lim_{z\rightarrow \infty}{}_1F_1(a,b,z)=    \frac{\Gamma(b)}{\Gamma(b-a)}(-z)^{-a}\Big(1+\mathcal{O}(1/z)\Big)+\frac{\Gamma(b)}{\Gamma(a)}e^z z^{a-b}  \Big(1+\mathcal{O}(1/z)\Big)\label{asymp}
    \end{gather}
    we see that the in-state wavefunction (\ref{NRQMinstate}) at large radial distances is composed of an incoming wave and a scattered wave
    \begin{gather}
        \lim_{|\vec{x}|\rightarrow \infty}\phi_{\text{in}}(\vec{x},\vec{p})=e^{-iEt}\Big(e^{i\vec{p}\.\vec{x}+i\gamma\log(|\vec{p}||\vec{x}|-\vec{p}\.\vec{x})}-\gamma\frac{\Gamma(1+i\gamma)}{\Gamma(1-i\gamma)}\frac{e^{i(|\vec{p}||\vec{x}|-\gamma\log(|\vec{p}||\vec{x}|-\vec{p}\.\vec{x}))}}{|\vec{p}||\vec{x}|-\vec{p}\.\vec{x}}\Big)\label{inexpansion}
    \end{gather}
 which is valid everywhere except the forward direction $\vec{x}\.\vec{p}=|\vec{x}||\vec{p}|$\footnote{This is so that $z$ in (\ref{asymp}) is large. The confluent hypergeometric function is still well-defined in the forward direction, where ${}_1F_1(a,b,0)=1$.} Thus, at early times, the in-state wavefunction appears like\footnote{Throughout this paper the phrase ``appears like'' serves as shorthand for considering a sharply peaked wave packet $\int \DD^3\vec{k} g(\vec{k})\,\, e^{-iE_kt} \phi_{\text{in}}(\vec{k},x)$ at early/late times and making a stationary-phase approximation when computing this integral in order to obtain a localized wave on a specifiable trajectory. An early/late-time stationary-phase approximation will pick out the first/second term on the RHS of (\ref{inexpansion}).} a classical particle travelling along a logarithmically corrected free-particle trajectory, and at asymptotically late times the in-state appears like a superposition of outgoing scattered waves. This justifies the name in-state. Similarly for the out-state. It is also worth noting that although each in/out state carries a momentum-like label $\vec{p}$, it never asymptotes to an eigenstate of the momentum operator with this label as its eigenvalue. Recall that the $S$-matrix is defined as the inner product of the in-state with the out-state (Eq. 3.2.1 of \cite{Weinberg:1995mt}). We can compute the exact $S$-matrix by taking the inner product of the in-state (\ref{NRQMinstate}) with the out-state (\ref{NRQMoutstate})
 \begin{align}
     A(\vec{p}_1,\vec{p}_2)&=\braket{\phi_{\text{out}}(\vec{p}_2)|\phi_{\text{in}}(\vec{p}_1)}\\
     &=\int\DD^3 x\,\, \phi^{\star}_{\text{out}}(\vec{x},\vec{p}_2)\phi_{\text{in}}(\vec{x},\vec{p}_1)\label{eq:innerprodNRQM}\\
     &=-\frac{i\pi}{2}\frac{e_1e_2}{|\vec{p}|^2}\delta(E_1-E_2)\frac{\Gamma(1+i\gamma)}{\Gamma(1-i\gamma)}\frac{1}{\sin^2(\theta/2)}e^{-i\gamma\log\sin^2(\theta/2)}\,\, ,\quad \theta\neq 0\label{exactNRQM}
 \end{align}
where $\theta$ is the scattering angle between the incoming and outgoing momenta. We evaluate the integral (\ref{eq:innerprodNRQM}) in appendix \ref{app:NRQM amp integral}. \par
Note that even though the last phase factor in (\ref{exactNRQM}) is a pure phase, it still leads to a measurable effect at the level of the cross section for the scattering of identical particles, due to an interference pattern between $t$- and $u$-channel exchanges; see Sect. 137 of \cite{Landau:1991wop}. This phase factor has been experimentally measured in the scattering of identical particles at low velocities \cite{chadwick}. \par
Recall that in addition to the $S$-matrix, it is common to work with an operator $\hat{S}$ defined by the condition that it yields the S-matrix when sandwiched between two free-particle states,  $\braket{\phi_{\text{out}}(p_2)|\phi_{\text{in}}(p_1)}\defined \braket{\vec{p}_2|\hat{S}|\vec{p}_1}$. With the exact in/out states in hand (\ref{NRQMinstate}, \ref{NRQMoutstate}), it is straightforward to write down the exact $\hat{S}$-operator
\begin{gather}
    \hat{S}=\int\DD^3p_2\int\DD^3p_1 \ket{\vec{p}_2}\braket{\phi_{\text{out}}(\vec{p}_2)|\phi_{\text{in}}(\vec{p}_1)}\bra{\vec{p}_1}\label{exactSmatrix}
\end{gather}
where $\ket{\vec{p}}$ are free-particle states, and (\ref{exactSmatrix}) straightforwardly satisfies the defining property of the $\hat{S}$-operator. Note that the exact $\hat{S}$-operator is IR-finite.
Note that the exact amplitude (\ref{exactNRQM}) has no IR cutoff or IR divergence, and similarly the Taylor expansion of (\ref{exactNRQM}) at each order in the coupling does not contain any cutoff. This provides part of the motivation for developing a perturbation theory for relativistic theories that similarly does not contain an IR cutoff at any order in the coupling expansion. Lastly, we note that the exact amplitude is far simpler than the usual perturbative expansion would suggest; for example, even the one-loop integral requires significant effort to evaluate. One can hope that a similar simplicity may hold for relativistic theories.
\subsection{On the definition of the Coulomb scattering amplitude}
Different textbooks report different expressions for the Coulomb scattering amplitude, which may give the impression that the probability amplitude is ambiguous. For example, Landau and Lifshitz \cite{Landau:1991wop} give the amplitude as (\ref{exactNRQM}), while Weinberg \cite{weinberg_2015} instead gives
\begin{gather}
-\frac{i\pi}{2} \frac{e_1 e_2}{|\vec{p}|^2} \delta(E_1 - E_2) \frac{\Gamma(1 + i\gamma)}{\Gamma(1 - i\gamma)} \frac{1}{\sin^2 (\theta/2)},
\end{gather}
which omits the phase factor $e^{-i\gamma\log\sin^2(\theta/2)}$. This discrepancy between textbooks may give the mistaken impression that the probability amplitude is ambiguous. To see why there is no ambiguity, we first need to recall an important theorem from non-relativistic scattering theory for short-range potentials (see, for example, Chapter 10 of \cite{Taylor:1972pty})
\begin{gather}
    f(E,\hat{p}\.\hat{p}')=-(2\pi)^2m\mathcal{T}(\vec{p}\rightarrow \vec{p}\,').\label{important theorem}
\end{gather}
The terms in (\ref{important theorem}) are defined as follows.  For a short-range potential, the in-state wavefunction has a large-distance expansion of the form
\begin{gather}
   \lim_{r\rightarrow\infty} \psi^{\t{in}}_{\vec{p}}(\vec{x})=\frac{1}{(2\pi)^{\frac{3}{2}}}\Big(e^{i\vec{p}\.\vec{x}}+f(E,\hat{p}\.\hat{x})\frac{e^{ipr}}{r}+\mathcal{O}(1/r^2)\Big)\label{large r}
\end{gather}
which defines the quantity $f(E,\hat{p}\.\hat{x})$. Next, we define \( \mathcal{T} \). For short-range potentials, the inner product of the in-state with the out-state takes the form
\begin{gather}
\braket{p'_{\text{out}}|p_{\text{in}}} = \delta^{(3)}(\vec{p} - \vec{p}\,') - 2\pi i \delta(E_p - E_{p'}) \mathcal{T}(\vec{p} \to \vec{p}\,'). \label{def of Tmatrix}
\end{gather}
On the left-hand side, we have the inner product of the in-state with the out-state, which in non-relativistic quantum mechanics is computed using \( \int d^3\vec{x} \). On the right-hand side, the first term represents a disconnected contribution, while the second term, after factoring out energy conservation, defines \( \mathcal{T}(\vec{p} \rightarrow \vec{p}\,') \).

We now have all the necessary definitions to understand theorem (\ref{important theorem}): the connected part of the inner product of the in-state with the out-state can be obtained directly from the large \( |\vec{x}| \) expansion of the in-state wavefunction, as specified in the theorem.

We can now explain why different textbooks report different results for the Coulomb amplitude. Unlike short-range potentials, the in/out states of the Coulomb potential do not allow for a large \( |\vec{x}| \) expansion of the form (\ref{large r}), meaning that theorem (\ref{important theorem}) does not apply—compare this to (\ref{inexpansion}) for the true large \( |\vec{x}| \) expansion. To our knowledge, no analog of this theorem has been established for long-range potentials. Nevertheless, textbooks take inspiration from theorem (\ref{important theorem}), which, as we have noted, does not hold in this case, to \textit{define} the Coulomb scattering amplitude as part of the coefficient of \( \frac{e^{ipr}}{r} \) in the large \( |\vec{x}| \) expansion of the in-state (\ref{inexpansion}). Different choices for what factor to extract lead to the different definitions for the Coulomb amplitude found in Weinberg and Landau \& Lifshitz.

However, there is no ambiguity in the probability amplitude. It is a fundamental axiom of quantum mechanics that probability amplitudes are given by inner products (Chapter 2 \cite{Weinberg:1995mt}). Since the exact in- and out-state wavefunctions are known, there is no need to rely on any theorem to compute the probability amplitude indirectly. The probability amplitude is unambiguously given by the \( \int d^3\vec{x} \) inner product of the in-state with the out-state, as in (\ref{exactNRQM}).  

\subsection{Origin of IR divergences}
Let us recall the derivation of the familiar form of the $\hat{S}$-operator in terms of Moller operators, Chapter 3 of \cite{Weinberg:1995mt}. The key assumption is that sharply peaked wave packets $g(\vec{k})$ of the exact in/out states $\psi_{\text{in/out}}(x,k)$, defined as eigenstates of the full Hamiltonian $H$, ``appear like'' free-particle states $\phi_0(x,k)$, defined as eigenstates of the free part of the Hamiltonian $H_0$, at early/late times, respectively\footnote{Throughout this paper the phrase ``appropriately,'' as used in the context of matching asymptotics, refers to the fact that in (\ref{keyassumption}) the $\phi_{0}$ states only need to match the early-time, large-distance asymptotics of $\psi_{\text{in}}$, whereas the asymptotics of $\phi_0$ and $\psi_{\text{in}}$ do not necessarily need to agree at late times. Similarly for the out-states.}
\begin{gather}
    \lim_{t\rightarrow -/+ \infty}\int\DD^3k \,\,e^{-iE_kt}g\big(\vec{k}\big)\big\vert \psi_{\text{in/out}}(k)\big\rangle=\lim_{t\rightarrow -/+ \infty}\int\DD^3k\,\, e^{-iE_kt}g\big(\vec{k}\big)\big\vert\phi_0(k)\big\rangle.\label{keyassumption}
\end{gather}
Presuming that (\ref{keyassumption}) is true, we rewrite the exponential on the LHS as $e^{-i H t}$ and that on the RHS as $e^{-iH_0t}$ to derive the Moller-operator expression for the exact in/out states
\begin{gather}
\big\vert \psi_{\text{in/out}}\big\rangle=\lim_{t\rightarrow -/+\infty}\Omega(t)    \ket{\phi_0(k)},\qquad \Omega(t)\defined e^{i Ht}e^{-iH_0t},\label{moller}
\end{gather}
which is understood to only yield meaningful results for smooth superpositions of energy eigenstates. From (\ref{moller}) the usual expression for the $\hat{S}$-operator in terms of Moller operators follows
\begin{gather}
\braket{\psi_{\text{out}}(p_2)|\psi_{\text{in}}(p_1)}\defined \braket{\phi_0(p_2)|\hat{S}|\phi_0(p_1)},\qquad \hat{S}= \Omega^{\dagger}(\infty)\Omega(-\infty).\label{smatrix}
\end{gather}
If the key assumption (\ref{keyassumption}) were true for the case of the NRQM Coulomb Hamiltonian when choosing $\ket{\phi_0}$ to be free-particle states and $H_0$ to be the free Hamiltonian, then the exact in-state would have the asymptotic form
\begin{gather}
    \lim_{|\vec{x}|\rightarrow \infty}\psi_{\text{in}}(\vec{x},\vec{p})=e^{i\vec{p}\.\vec{x}}+\frac{e^{i|\vec{p}||\vec{x}|}}{|\vec{x}|}f_{\vec{p}\,}(\hat{x})+\mathcal{O}(1/|\vec{x}|^2)\label{asymptotic form}
\end{gather}
The early-time stationary-phase analysis would pick out the first term on the RHS of (\ref{asymptotic form}). Thus, (\ref{asymptotic form}) is a direct consistency condition for the exact in-state to appear like a free particle at asymptotically early times. However, we see from the expansion of the exact in-state (\ref{inexpansion}) that (\ref{asymptotic form}) is not true. Consequently, the Moller operator in the form (\ref{moller}), acting on a free-particle state, does not asymptote to the exact in-state; using the exact solution (\ref{NRQMinstate}), one can verify that it does not converge to a well-defined limit at all. We conclude that IR divergences arise in standard perturbation theory because the key assumption (\ref{keyassumption}) is not satisfied. Two possible fixes are either to modify the Moller operators while continuing to act on free-particle states, which is the Dollard-Faddeev-Kulish approach that we will review shortly, or, as we pursue in this paper, to use states $\ket{\phi}$ which satisfy the key assumption (\ref{keyassumption}) with a minor modification to what we call $H_0$.
\subsection*{Dollard's Moller operators}
Dollard \cite{Dollard1964AsymptoticCA} remedies the situation in NRQM by finding Moller operators 
\begin{gather}
    \Omega^{D}_{\mp}=\lim_{t\rightarrow \mp\infty}e^{iHt}\exp\Big\lbrace-i\Big(H_0t\pm \frac{e_1e_2}{4\pi} \sqrt{\frac{m}{2}}\frac{1}{\sqrt{H_0}}\log(4H_0|t|)\Big)\Big\rbrace\label{dollardmoller}
\end{gather}
which correctly give the exact in/out states when acting on free-particle states $\braket{\vec{x}|\vec{p}}=e^{-i\vec{p}\.\vec{x}}$ (see (5) of \cite{barrachina1989scattering} and (54) of \cite{Dollard1964AsymptoticCA})
\begin{gather}
\braket{\vec{x}|\hat{\Omega}^D_{\mp}|\vec{p}}=\Gamma(1\pm i\gamma)e^{-\frac{\pi\gamma}{2}}e^{-iEt+i\vec{p}\.\vec{x}}{}_1F_1\Big(\mp i\gamma,1,\pm i|\vec{p}||\vec{x}|-i\vec{p}\.\vec{x}\Big).\label{Dollardproject}
\end{gather}
In order to compare the perturbation theory developed in this paper to the Faddeev-Kulish (FK) \cite{Kulish:1970ut} approach we note that FK give a relativistic generalization of Dollard's Moller operators
\begin{gather}
\Omega^{\text{FK}}_{\mp}=\lim_{t\rightarrow \mp \infty}\Omega(t) e^{-i\Phi (t)}e^{R(t)}\label{Roperator}\\
    \Phi(t) = -\frac{\alpha}{2} \int \frac{d^3p}{(2\pi)^3} \frac{d^3q}{(2\pi)^3} : \hat{\rho}(p) \hat{\rho}(q) : \frac{p \cdot q}{\sqrt{(p \cdot q)^2 - m^4}} \text{sign}(t)\ln \frac{|t|}{t_0}\label{FKcoulomb phase operator}
\end{gather}
where $\hat{\rho}$ are charge operators, $\Omega(t)$ denotes the standard Moller operator (\ref{moller}), and $R(t)$ is a term associated with real-photon production, a process that we do not consider in this paper; see the discussion below (\ref{eq:Weinbergexponential}). There have been very few explicitly calculated amplitudes using this relativistic Coulomb-phase operator (\ref{FKcoulomb phase operator}). This is likely because it is not immediately apparent how to implement this operator in practical amplitude calculations or how energy conservation will follow.
 \subsection{Runge-Lenz symmetry}\label{sect:reviewrunge}
 \footnote{This section is outside the main focus of this article and can be skipped on a first reading.} The Coulomb Hamiltonian (\ref{eq:CoulombHamiltonian}) possesses an additional symmetry beyond rotational invariance,
 \begin{gather}
     [H,\vec{A}]=0\\
     \vec{A}=\frac{1}{2m}(\vec{p}\times \vec{L}-\vec{L}\times \vec{p})-\frac{e_1e_2}{4\pi} \hat{x}
 \end{gather}
 where $\vec{A}$ is referred to as the Runge-Lenz vector. On the space of positive energy states (unbound scattering states) the rotation and Runge-Lenz operators close into a $SO(1,3)$ algebra, and on the negative energy (bound state) wavefunctions the algebra closes into $SO(4)$, see \cite{Zwanziger} for details.  It was first noticed in \cite{Zwanziger} that the in/out states (\ref{NRQMinstate}, \ref{NRQMoutstate}) are eigenstates of the Runge-Lenz operator in the direction of their momentum
\begin{gather}
  \frac{\vec{k}}{|\vec{k}|}\cdot \Big[(\hat{p}\times \hat{L})-i \hat{p}+m\frac{e_1e_2}{4\pi}\frac{\vec{x}}{r}\Big]\phi_{\frac{\text{in}}{\text{out}}}(\vec{x},\vec{k})=\Big(\mp\frac{m}{k}\frac{e_1e_2}{4\pi}+i \Big)\phi_{\frac{\text{in}}{\text{out}}}(\vec{x},\vec{k})
\end{gather}
where the terms in square brackets are the familiar vector differential operators from quantum mechanics, and the minus/plus signs for the eigenvalue on the RHS are for the in/out states, respectively. This equation is the harbinger of the fact that the in/out states have conformal dimension $(1+i\gamma)$ under the action of the $SO(1,3)$ Runge-Lenz symmetry. There have been many studies of the Runge-Lenz symmetry of the non-relativistic Coulomb Hamiltonian \cite{bander1966group,bander1966group2,Zwanziger,wu1987group,stahlhofen1997algebraic,kerimov2005models}, and several suggestions for how this symmetry may be extended to relativistic settings \cite{dahl1997physical,Caron-Huot:2014gia,Alvarez-Jimenez:2018lff,barut1973hydrogen,barut1971nonrelativistic,barut1974relativistic}. The non-relativistic studies have focused on the implications of this symmetry for the partial-wave expansions, where it is known that the states furnish a principal-series representation of $SO(1,3)$. Below we point out an observation that follows straightforwardly from these studies, namely that the exact amplitude takes the form of a $2$-point correlation function of operators in a $2$D CFT on a sphere. Let us parameterize the unit 3-momenta using stereographic coordinates:
\begin{gather}
    \hat{k}=\frac{1}{1+|z|^2}\Big(z+\bar{z},i(\bar{z}-z),1-|z|^2\Big)\\
    z=e^{i\phi}\tan\frac{\theta}{2}
\end{gather}
With this parametrization, the exact scattering amplitude (\ref{exactNRQM}) becomes
     \begin{gather}
     A(p_1,p_2)=-\frac{i\pi}{2}\frac{e_1e_2}{|\vec{p}|^2}\Omega(z_1)^{\Delta}\Omega(z_2)^{\Delta}\delta(E_1-E_2)\frac{\Gamma(1+i\gamma)}{\Gamma(1-i\gamma)}\frac{1}{|z_1-z_2|^{2\Delta}},\qquad \Delta=1+i\gamma\label{eq:conformalexact}
 \end{gather}
 where $\Omega(z)=(1+|z|^2)$ can be interpreted as a Weyl-rescaling factor which arises from conformally mapping the plane to the sphere. From (\ref{eq:conformalexact}) the reader can recognize that the exact NRQM scattering amplitude takes the form of a $2$-point function of scalar conformal primary operators on a sphere on the principal series. Note that the $SO(1,3)$ conformal transformations act only on the $z$ coordinates, $z\mapsto \frac{az+b}{cz+d}$, and not on the particle's energy, as the Runge-Lenz vector commutes with the Hamiltonian and hence does not change the particle's energy. Note that the energy-conserving delta function also ensures that the conformal dimensions of the in/out ``operators'' are the same, as they must be in a 2-point CFT correlator. We find in section \ref{sect:runge lenz} that the LO relativistic amplitudes in this paper also enjoy this symmetry.
\section{Defining the semiclassical scattering amplitude}\label{sect:setup}
In this section we define the scattering amplitude, which we will compute in the rest of this paper, to be the inner product of two solutions to the Klein-Gordon equation in the Coulomb background of a source charge.\\
Consider the equation of motion (EQM) for a scalar field $\psi$ on a fixed background gauge field $A^{\mu}$,
\begin{equation}
    (\Box +m^2+2ie_1A^{\mu}\d_{\mu})\psi= (e_1^2A^{\mu}A_{\mu})\psi\label{eq:intmotion}
\end{equation}
We can introduce an inner product\footnote{This is not strictly an inner product as it does not satisfy the positive definiteness condition for negative energy solutions. We will only be computing inner products amongst positive energy solutions.} on the space of solutions to this equation,
\begin{align}
\braket{\psi_1|\psi_2}_{\text{SQED}}&=i\int_{\Sigma}\DD^3\Sigma_{\mu}J^{\mu}\\
    &=i\int_{\Sigma}\DD^3\Sigma_{\mu}\Bigg(\psi_1^{\star}(x)\d^{\mu}\psi_2(x)-\psi_2(x)\d^{\mu}\psi^{\star}_1(x)+2ie_1 \psi_1^{\star}\psi_2 A^{\mu}\Bigg)\label{eq:defJ}
\end{align}
where $\Sigma$ is an arbitrary Cauchy surface and $\DD^3\Sigma_{\mu}$ is the volume element which is normal to the surface. We add a subscript SQED to the inner-product brackets to remind us to use the scalar-QED inner product. This inner product is independent of the Cauchy surface\footnote{So long as the current falls off sufficiently fast at the boundaries of the Cauchy surface.} because the current is divergence-free, $\d_{\mu}J^{\mu}=0$, on the support of the equations of motion (\ref{eq:intmotion}) when working in Lorenz gauge $\d_{\mu}A^{\mu}=0$. Furthermore, the inner product is Lorentz invariant because $J^{\mu}$ is a Lorentz vector. The $i$ prefactor is necessary to ensure that the inner product is Hermitian. Let us note that the same inner product can be used for the classical solution space of
\begin{equation}
    (\Box +m^2+2ie_1A^{\mu}\d_{\mu})\phi= 0\label{eq:freemotion}
\end{equation}
as the reader can verify that the same $J^{\mu}$ in (\ref{eq:defJ}) is divergence-free on the support of (\ref{eq:freemotion}). Throughout this paper we reserve $\psi$ for solutions to the full EQM (\ref{eq:intmotion}) and $\phi$ for solutions to the ``semi-free'' EQM (\ref{eq:freemotion}).\\
In this paper we will fix the background gauge field to be the field produced by a charged particle moving with constant four-velocity\footnote{Throughout we work in the mostly minus metric signature convention $\eta=(+---)$.} 
\begin{equation}
    A^{\mu}(x)=\frac{e_s}{4\pi}\frac{u_s^{\mu}}{\sqrt{(u_s\.x)^2-x^2}}\label{eq:gauge field}
\end{equation}
where $u^{\mu}_s=\frac{p^{\mu}_s}{m_s}$ is the four-velocity of the source particle, and the gauge field (\ref{eq:gauge field}) satisfies the Lorenz gauge condition $\d_{\mu}A^{\mu}=0$. Note that at this point we are deviating from the quantum theory, where the gauge field responds to a source charge via the Feynman propagator rather than the retarded propagator which we used to construct (\ref{eq:gauge field}). The Feynman propagator would introduce an additional term in (\ref{eq:gauge field}), identical in form to the real-emission term in (\ref{eq:Weinbergexponential}). Indeed, if we used the retarded propagator for the photon when evaluating the integral that led to the exponential factor (\ref{eq:Weinbergexponentiation}) (see Eq. 13.2.3 of \cite{Weinberg:1995mt} for the integral), then the emission term $R$ would not be present. Thus, by using the retarded gauge field (\ref{eq:gauge field}), we are able to sidestep the issue of infrared divergences associated with real-photon emission, thereby allowing us to isolate the focus of this paper: the Coulomb-phase divergence. \\
Let us now presuppose that we have two solutions $\psi_{\text{in}}(p_1,x),\psi_{\text{out}}(p_2,x)$ to the full equations of motion (\ref{eq:intmotion}) in the background field (\ref{eq:gauge field}) which respectively at asymptotically early and late times appear approximately like particles travelling along classical trajectories one would associate with four-momenta $p_{1}^{\mu}$ and $p_{2}^{\mu}$. We then define the scattering amplitude to be the inner product of these two solutions
\begin{gather}
    \mathcal{A}(p_1,p_2,u_s)\defined\braket{\psi_{\text{out}}(p_2,u_s)|\psi_{\text{in}}(p_1,u_s)}_{\text{SQED}}.\label{eq:amplitude}
\end{gather}
This will be the object which we perturbatively construct in the rest of this paper. This is one of the simplest yet prototypical amplitudes which exhibits the Coulomb-phase divergence if one uses standard perturbation-theory techniques \cite{kang1962higher}. This amplitude can also be computed explicitly by non-perturbative means; see (\ref{eq:exactamp}) for the exact IR-finite amplitude and appendix \ref{subapp:exactinout} for a derivation. We note that the definition (\ref{eq:amplitude}) of the semiclassical scattering amplitude also coincides with the definition of the ``tree-level'' scattering amplitude in the perturbiner approach to scattering on a fixed background (cf. \cite{DeWitt,Arefeva:1974jv,ABBOTT1983372,Jevicki,Rosly_1997,rosly1997gravitational,selivanov1999postclassicism,Mizera_2018,Garozzo_2019,ilderton2023scattering,Adamo:2021rfq}).
\section{Perturbation theory}\label{sect:IR finite perturbation theory}
\subsection{Relativistic Coulomb wavefunctions}
We will perturbatively solve the full equations of motion (\ref{eq:intmotion}) by perturbing about the solution to the equations of motion in the absence of the $A^{\mu}A_{\mu}\phi$ term, see (\ref{eq:freemotion}), which we will refer to as the semi-free equations of motion. In order to derive the solution to the semi-free equations of motion, we first solve them in the rest frame of the source particle $u_s^{\mu}=(1,\vec{0})=\delta^{\mu}_0$, so that the equations take the form of the NRQM equations of motion
\begin{gather}
    (\Box+m^2)\phi(x)=-\frac{2ie_1e_s}{4\pi r}\d_{t}\phi\,.
\end{gather}
We can then take the standard results from NRQM textbooks to find the solution
\begin{align}
\phi_{\text{in}}\Big(x,p,u_s^{\mu}=\delta^{\mu}_0\Big)&=\Gamma(1+i\gamma)e^{-\frac{\pi\gamma}{2}}e^{-iEt+i\vec{p}\.\vec{x}}{}_1F_1\Big(-i\gamma,1,i(|\vec{p}||\vec{x}|-\vec{p}\.\vec{x})\Big)\label{restframeinstate}\\
    \phi_{\text{out}}\Big(x,p,u_s^{\mu}=\delta^{\mu}_0\Big)&=\Gamma(1-i\gamma)e^{-\frac{\pi\gamma}{2}}e^{-iEt+i\vec{p}\.\vec{x}}{}_1F_1\Big(i\gamma,1,-i|\vec{p}||\vec{x}|-i\vec{p}\.\vec{x}\Big)\label{restframeoutstate}
    \end{align}
    \begin{equation}
        \gamma(p,u_s=\delta^{\mu}_0)\defined \frac{e_1e_s E}{4\pi |\vec{p}|},\qquad E=\sqrt{|\vec{p}|^2+m^2}
    \end{equation}
The key difference between the NRQM wavefunctions (\ref{NRQMinstate}) and the relativistic wavefunctions (\ref{restframeinstate}) is the energy dependence of the conformal dimension, $\frac{e_1e_s}{4\pi}\frac{m}{|\vec{p}|}\rightarrow \frac{e_1e_s}{4\pi}\frac{E}{|\vec{p}|}$. We can then ``covariantize'' these results to find the solution in an arbitrary Lorentz frame
    \begin{align*}
     &\phi_{\text{in}}(x,p,u_s)=\Gamma(1+i\gamma_{ps})e^{-\frac{\pi\gamma_{ps}}{2}}e^{-iE t+i\vec{p}\.\vec{x}}\\
     &{}_1F_1\Bigg(-i\gamma_{ps},1,i \Big(\sqrt{(p\.u_s)^2-p^2}\sqrt{(u_s\.x)^2-x^2}-(p\.u_s)(u_s\.x) +(p\.x)\Big)\Bigg)\numberthis \label{eq:instate}\\
       &\phi_{\text{out}}(x,p,u_s)=\Gamma(1-i\gamma_{ps})e^{-\frac{\pi\gamma_{ps}}{2}}e^{-iE t+i\vec{p}\.\vec{x}}\\
     &{}_1F_1\Bigg(i\gamma_{ps},1,i \Big(-\sqrt{(p\.u_s)^2-p^2}\sqrt{(u_s\.x)^2-x^2}-(p\.u_s)(u_s\.x) +(p\.x)\Big)\Bigg)\numberthis \label{eq:outstate}
     \end{align*}
     where
     \begin{align*}
     &\gamma_{ps}=\gamma(p,u_s)=\frac{e_1e_s}{4\pi }\frac{p\.u_s}{\sqrt{(p\.u_s)^2-p^2}}=\frac{e_1e_s}{4\pi}\frac{1}{\beta_{ps}},\qquad E=\sqrt{|\vec{p}|^2+m^2}\numberthis
\end{align*}
which indeed satisfy
\begin{gather}
    \Big(\Box +m^2+2i\frac{e_1e_s}{4\pi}\frac{u_s^{\mu}}{\sqrt{(u_s\.x)^2-x^2}}\d_{\mu}\Big)\phi_{\text{in/out}}= 0.
\end{gather}
Note that in all these expressions the momenta are on shell, $p^2=m^2$; however, we have written them in a form that will allow us to take them off shell in the following sections. As shown in appendix \ref{subapp:ortho}, these wavefunctions (\ref{eq:instate}, \ref{eq:outstate}) are orthonormal with respect to the SQED inner product
\begin{gather}
    \braket{\phi_{\text{in}}(p_1)|\phi_{\text{in}}(p_2)}_{\text{SQED}}=\braket{\phi_{\text{out}}(p_1)|\phi_{\text{out}}(p_2)}_{\text{SQED}}=(2\pi)^3 2E_1\delta^3(\vec{p}_1-\vec{p}_2).\label{orthonormal}
\end{gather}
\subsection{Relativistic Coulomb Green's function} 
We would like to solve the differential equation
\begin{gather}
    (\Box +m^2+2ie_1A^{\mu}\d_{\mu})\psi(x)=V(x)\psi(x)\label{fulleq}\\
    V(x)=e_1^2A^{\mu}(x)A_{\mu}(x)
\end{gather}
We can obtain a formal solution to (\ref{fulleq}) by multiplying through by the inverse of the differential operator occurring on the LHS of (\ref{fulleq}) to obtain a Lippmann-Schwinger type equation for the solution 
\begin{gather}
    \psi(x)=\phi(x)+\int\DD^4 y \,\, G(x,y)V(y)\psi(y)\label{eq:LS}
\end{gather}
where $\phi(x)$ is in the kernel of $(\Box +m^2+2ie_1A^{\mu}\d_{\mu})$, and $G(x,y)$ denotes the inverse of this operator
\begin{gather}
    (\Box +m^2+2ie_1A^{\mu}\d_{\mu})\phi(x)=0\\
    (\Box +m^2+2ie_1A^{\mu}\d_{\mu})_xG(x,y)=\delta^{4}(x-y).\label{greenseq}
\end{gather}
Note that the full solution $\psi(x)$ occurs on both sides of (\ref{eq:LS}), hence only providing a formal solution. However, we can iteratively plug this equation back into itself to obtain a perturbative expansion for the wavefunction,
\begin{gather}
    \psi^{\text{in/out}}(x)=\phi^{\text{in/out}}(x)+\int\DD^4 y \,\, G^{+/-}(x,y)V(y)\phi^{\text{in/out}}(y)+\ldots\label{refer}
\end{gather}
where the superscript $+/-$ on the Green's function indicates that the boundary conditions imposed on the Green's function depend on whether we are constructing the in- or out-state. Note that in standard perturbation theory one takes the first term on the RHS of (\ref{refer}) to be a plane wave, whereas our experience with the exact NRQM in/out states (\ref{inexpansion}) indicates that this is not a good approximation to the asymptotics of the exact wavefunction. \\
We now turn to constructing the Coulomb Green's function \cite{1964JMP.....5..591H}. We can use a spectral decomposition to write the Green's function as a sum over its eigenvectors divided by their eigenvalues. To this end we will need the spectrum of the operator $(\Box +m^2+2ie_1A^{\mu}\d_{\mu})$. We can construct the eigenvectors by taking our semi-free solutions (\ref{eq:instate},\ref{eq:outstate}) off shell
\begin{gather}
   (\Box +m^2+2ie_1A^{\mu}\d_{\mu})_x\chi(x,p,u_s)_{\text{in/out}}=(-p^2+m^2)\chi(x,p,u_s)_{\text{in/out}} \label{eigenvaluecontiniuum}
\end{gather}
where $\chi(x,p,u_s)_{\text{in/out}}$ are verbatim the Coulomb wavefunctions in (\ref{eq:instate},\ref{eq:outstate}), except that we no longer impose\footnote{Of course, the source particle's four-momentum is still on shell, $u_s^2=1$.} $p^{0}=\sqrt{|\vec{p}|^2+m^2}$. The Green's function is then
\begin{gather}
    G^{\pm}(x,y)=-\int\frac{\DD^4p}{(2\pi)^4} \,\, \frac{\chi_{\text{in/out}}(x,p,u_s)\chi_{\text{in/out}}^{\star}(y,p,u_s)}{p^2-m^2\pm i\e}+\text{Bound states}\label{eq:Greensfunction}
\end{gather}
where the bound-state contribution will not feature significantly at the orders to which we compute amplitudes in this paper, so we relegate the explicit expression for these bound-state terms to appendix \ref{app:boundstates}; see in particular (\ref{boundstate contibution}). We are free to use either the in-state or out-state wavefunctions to express the Green's function, as both sets form a complete basis, and we will make convenient use of this freedom later when calculating amplitudes. Appendix \ref{app:Greensfunction} examines the Green's function (\ref{eq:Greensfunction}) in more detail and verifies that it satisfies all of the required properties. We give the off-shell bound-state contribution to the Green's function in appendix \ref{app:boundstates} and demonstrate in appendix \ref{app:proveGreens} that (\ref{eq:Greensfunction}) satisfies the Green's function equation (\ref{greenseq}). For an in-state/out-state, we use the retarded/advanced $G^{+/-}$ propagator and place both $p^2=m^2$ on-shell poles below/above the real $p^0$ axis, as usual. As a point of familiarity, note that if we take the coupling to zero we obtain the familiar Green's function of a free scalar theory
\begin{gather}
    \lim_{e_s\rightarrow 0}G^{\pm}(x,y)=-\int\frac{\DD^4p}{(2\pi)^4} \frac{e^{-ip\.(x-y)}}{p^2-m^2\pm i\e},
\end{gather}
where we used that ${}_1F_1(0,b,z)=1$, and that the bound states are proportional to the coupling. The prefactors $\Gamma(1\pm i\gamma)$ in $\chi_{\text{in/out}}$ contain simple poles in the complex $p^0$ plane; see (\ref{eq:instate}, \ref{eq:outstate}). Because typical amplitude calculations involve deforming the $p^0$ integration contour in (\ref{eq:Greensfunction}) to pick up the on-shell poles in the complex $p^0$ plane, we analyze the pole structure of the Green's function in appendix \ref{app:boundstatecancellation}. There, we demonstrate that these poles are cancelled by part of the bound-state contribution to the Green's function, as they must be when imposing retarded/advanced boundary conditions. The upshot of the analysis in appendix \ref{app:boundstatecancellation} is that one can work with the following purely on-shell expression for the Green's functions
\begin{align*}
    G^{\pm}(x,y)&=\pm i\, \theta\Big(\pm(x^0-y^0)\Big)\Bigg\lbrace
    \sum_{l=0}^{\infty}\sum_{n=l+1}^{\infty}\sum_{m=-l}^lE_n\phi_{nlm}(x)\phi^{\star}_{nlm}(y)\\
    &+\int\frac{\DD^3 p}{(2\pi)^3}\frac{1}{2E_p} \Big(\phi_{\text{in/out}}(x,p,u_s)\phi_{\text{in/out}}^{\star}(y,p,u_s)-(E_p\rightarrow -E_p)\Big)
    \Bigg\rbrace \numberthis\label{eq:onshellgreens}
\end{align*}
where the first term in (\ref{eq:onshellgreens}) is a sum over the on-shell bound states, the explicit expression for which is given in appendix \ref{app:boundstates}. The second term in (\ref{eq:onshellgreens}) is an integral over the in/out wavefunctions (\ref{eq:instate}, \ref{eq:outstate}). When one takes the coupling to zero, the Coulomb wavefunctions become plane waves and we recover the standard on-shell expression for the Green's functions of a massive scalar. \par
There have been several studies of the non-relativistic (NR) Coulomb Green's function \cite{Schwinger:1964zzb,osti_4377409,revisitcoulomb} (and references therein), where the majority of the literature has focused on expressing the continuum contribution to the Green's function in terms of partial waves, whereas we have sought an expression involving an integral over the scattering states, namely the relativistic version of \cite{mano1963eigenfunction,mano1964representations}. We note the very useful paper by Hostler \cite{1964JMP.....5..591H}, who considered the relativistic Coulomb Green's function; however, the expressions given therein are the full position-space expressions for the Green's functions, which are not appropriate for our purposes of calculating scattering amplitudes.

\subsection{Summary and relation to Faddeev-Kulish}\label{sect:summary}
Let us recap the above proposed perturbation theory. By iterating the Lippmann-Schwinger equation (\ref{eq:LS}) with the starting input solutions of either (\ref{eq:instate}) or (\ref{eq:outstate}) for $\phi(x)$, and using (\ref{eq:onshellgreens}) for the Green's function,  we obtain perturbative solutions to the full classical equations of motion (\ref{eq:intmotion}). We can then take the SQED inner product (\ref{eq:defJ}) of the resultant wavefunctions to compute the scattering amplitude (\ref{eq:amplitude}). On the level of Feynman diagrams, we have reorganized the perturbation theory according to the number of $A^2\phi^2$ quartic vertices, as the Coulomb wavefunctions contain all of the information about the sum over all $A\phi^2$ photon exchanges between the source and matter particle. \par 
The fact that the solutions we are perturbing about, $\phi_{\text{in/out}}(x,p,u_s)$, contain all of the dynamics of the $\frac{1}{r}$ potential leads us to expect that the resultant perturbation theory does not contain any infrared divergences. More to the point, the perturbing potential $V(x)=e_1^2A^{\mu}A_{\mu}$ falls off as $\sim \frac{1}{r^2}$ at large distances, thus giving us the expectation that at large distances the asymptotics of the exact in/out states $\psi_{\text{in/out}}$ appropriately match, in the precise sense of (\ref{keyassumption}), the asymptotics of the input wavefunctions $\phi_{\text{in/out}}$\footnote{Except potentially in the exactly forward direction at zero impact parameter.} As infrared divergences arise due to a mismatch between the asymptotics of the input solution and the exact solution, this further motivates our expectation that the perturbation theory is IR-finite. These heuristic arguments aside, we explicitly verify in appendix \ref{subapp:verifyasymp} that the Coulomb wavefunctions (\ref{eq:instate}, \ref{eq:outstate}) satisfy the key assumption (\ref{keyassumption}), namely that the early-time asymptotics of the in-state Coulomb wavefunction $\phi_{\text{in}}$ appropriately match the early-time asymptotics of the exact in-wavefunction $\psi_{\text{in}}$. Similarly for the out-state wavefunctions. \par
Comparing this Coulomb-wavefunction perturbation theory to the Faddeev-Kulish approach is challenging, as there are few examples in the literature of using the Coulomb-phase operator (\ref{FKcoulomb phase operator}) to compute a scattering amplitude. For the case of scattering on a fixed background, without soft-photon production\footnote{So that we can ignore the $R$ operator in (\ref{Roperator}), as this paper only addresses the Coulomb-phase divergence.}, we expect the results to agree for the following reason.
 In the non-relativistic case, Dollard's Moller operator $\Omega^{D}_{\text{in/out}}$ (\ref{dollardmoller}) converges to an operator that projects free-particle plane waves $\ket{\vec{p}}$ onto the NRQM Coulomb wavefunctions $\phi_{\text{in/out}}$ (\ref{NRQMinstate}, \ref{NRQMoutstate}), 
\begin{gather}
    \hat{\Omega}^D_{\text{in/out}}=\int\DD^3 p \,\,\big|\phi_{\text{in/out}}(p)\big\rangle\big\langle \vec{p}\,\big\vert\label{Dollardprojection}
\end{gather}
(see the discussion around Eq. (54) in \cite{Dollard1964AsymptoticCA} and Eq. (94) in \cite{dollard1971quantum} for more details). As the FK Coulomb-phase operator (\ref{FKcoulomb phase operator}) is a relativistic generalization of Dollard's Moller operator, we expect that the initial stages of the FK evolution will project free-particle states onto the relativistic generalization of the NRQM Coulomb wavefunctions, namely the wavefunctions (\ref{eq:instate}, \ref{eq:outstate}). To be more precise, the FK Moller operators are
\begin{gather}
    \Omega^{\text{FK}}_{-/+}=\lim_{t\rightarrow -/+\infty}e^{iHt}e^{-iH_{0}t}e^{-i\Phi(t)}
\end{gather}
where $H,H_0$ are the full and free Hamiltonians, respectively, for a KG particle in a fixed background Coulomb field, and $\Phi$ is
\begin{equation}
      \Phi(t) = -\frac{\alpha}{2} \int \frac{d^3p}{(2\pi)^3}   \hat{\rho}(p)  \frac{p \cdot u_s}{\sqrt{(p \cdot u_s)^2 - m^2}} \text{sign}(t)\ln \frac{|t|}{t_0}
\end{equation}
which is the appropriate modification of the $\Phi$ in (\ref{FKcoulomb phase operator}) for acting on a single particle state in a fixed Coulomb background sourced by a charged particle with four-velocity $u_s^{\mu}$. 
If we let $H_r$ denote the Hamiltonian with all terms except for the quartic vertex $A^2\phi^2$,  then we can write the FK Moller operator acting on a free particle state $\ket{p}$ as
\begin{align}
      \Omega^{\text{FK}}_{-/+}\ket{p}=\lim_{t\rightarrow -/+ \infty}e^{iHt}e^{-iH_rt}e^{iH_rt}e^{-iH_{0}t}e^{-i\Phi(t)} \ket{p}\label{3exp}\\
      =\lim_{t\rightarrow -/+ \infty}e^{iHt}e^{-iH_rt}\ket{\phi_{\text{in/out}}(p)}\label{eq:nonrigor}
\end{align}
where in (\ref{3exp}) we have strategically placed the identity operator $\mathds{1}=e^{-iH_rt}e^{iH_rt}$, and here $\phi_{\text{in/out}}$ are the relativistic Coulomb wavefunctions (\ref{eq:instate}, \ref{eq:outstate}), which are eigenstates of $H_r$. In going to the last line we used that the last three exponentials in (\ref{3exp}) are the relativistic generalization of Dollard's Moller operator for scattering on the $\frac{1}{r}$ potential, without the $\frac{1}{r^2}$ quartic vertex, and thus we expect this operator to implement the relativistic version of the projection (\ref{Dollardprojection}). We make no claim to the rigor of the final step (\ref{eq:nonrigor}), as one would need to verify that the presence of the first two exponentials on the RHS of (\ref{3exp}) do not interfere with the proof \cite{dollard1971quantum} that the last three exponentials act as a projection operator from free states onto Coulomb states. Here we only intend to indicate where one might look for a direct proof of the equivalence of the two methods.
\section{Infrared-finite scattering amplitudes}\label{sect.Amplitudes}
In this section we will compute the scattering amplitude up to next-to-leading order (NLO)
\begin{equation}
\begin{aligned}
    \mathcal{A}(p_1,p_2,u_s)&\defined\braket{\psi_{\text{out}}(p_2,u_s)|\psi_{\text{in}}(p_1,u_s)}_{\text{SQED}}\\
    &=\braket{\phi_{\text{out}}(p_2,u_s)|\phi_{\text{in}}(p_1,u_s)}_{\text{SQED}}\\
&\quad+\braket{\phi_{\text{out}}(p_2,u_s)|\hat{G}^{+}\hat{V}|\phi_{\text{in}}(p_1,u_s)}_{\text{SQED}}\\
&\quad+\braket{\phi_{\text{out}}(p_2,u_s)|\hat{V}^{\dagger}(\hat{G}^{-})^{\dagger}|\phi_{\text{in}}(p_1,u_s)}_{\text{SQED}}+\ldots
\end{aligned}
\label{LO and NLO def}
\end{equation}
where we refer to the first term in the expansion on the RHS of (\ref{LO and NLO def}) as the LO amplitude and the remaining two terms as the NLO amplitude.
\subsection{LO amplitude}
The leading order amplitude is
\begin{align}
  \mathcal{A}_0(p_2,p_1;u_s)&=\braket{\phi_{\text{out}}(p_2,u_s)|\phi_{\text{in}}(p_1,u_s)}_{\text{SQED}}\\
 &=i\int_{\Sigma}\DD^3\Sigma_{\mu}\, \Big[\phi_{\text{out}}^{\star}(x,p_2)D^{\mu}\phi_{\text{in}}(x,p_1)-\phi_{\text{in}}(x,p_1)\Big(D^{\mu}\phi_{\text{out}}(x,p_2)\Big)^{\star}\Big]\label{eq:LO}
\end{align}
where $\phi_{\text{in/out}}$ are given in (\ref{eq:instate}, \ref{eq:outstate}), $D^{\mu}=\d^{\mu}+ie_1A^{\mu}$, and for the sake of brevity we omit the dependence of all terms in (\ref{eq:LO}) on the source particle's four-velocity $u_s^{\mu}$. In appendix \ref{app:LOamp} we evaluate this integral. There, in order to simplify the calculation, we exploit the Lorentz invariance of the LO amplitude to first evaluate the amplitude in the rest frame of the source particle, $u_s^{\mu}=\delta^{\mu}_0$. In order to compare the amplitude to the exact non-perturbative amplitude in section \ref{sect:exactresult}, we first express the result in terms of its Legendre-polynomial $P_{l}$ expansion
\begin{align}
\mathcal{A}_0\Big(p_2,p_1;u^{\mu}_s=\delta^{\mu}_0\Big)&= \frac{4\pi^2 \delta(E_1-E_2)}{|\vec{p}_1|}\sum_{l}(2l+1)\frac{\Gamma(1+l+i\gamma_{1s})}{\Gamma(1+l-i\gamma_{1s})}P_{l}(\hat{p}_1\.\hat{p}_2)\label{tree:spherical}\\
  &=-i\gamma\frac{4\pi^2}{|\vec{p}_1|} \delta(E_1-E_2) \frac{\Gamma(1+i\gamma)}{\Gamma(1-i\gamma)}\Bigg(\frac{2|\vec{p}_1|}{|\vec{p}_1-\vec{p}_2|}\Bigg)^{2+2i\gamma}\label{treerestfframe}
\end{align}
We note that the sum indicated in (\ref{tree:spherical}) does not converge, and the equality to (\ref{treerestfframe}) should be understood as an equality of distributions that behave the same under an integral sign \cite{Taylor}. This can be verified by projecting both (\ref{tree:spherical},\ref{treerestfframe}) onto $P_{l'}(\cos\theta)$ and using Rodrigues' formula, while giving $\gamma$ a positive infinitesimal imaginary part at intermediate stages. This sum has also been verified in alternative ways in \cite{Lin_2000,li2021eliminating,mott1932polarisation}.
We can then ``covariantize'' the result (\ref{treerestfframe}) to an arbitrary Lorentz frame
\begin{align*}
    &\mathcal{A}_0\Big(p_2,p_1;u_s\Big)\\
    &=i\pi\, e_1e_s\delta\Big(u_s\.(p_1-p_2)\Big)\frac{\Gamma(1+i\gamma_{1s})}{\Gamma(1-i\gamma_{1s})}\frac{p_1\.u_s}{m^2-(p_1\.u_s)^2}\Bigg(\frac{4\big(m^2-(p_1\.u_s)^2\big)}{(p_1-p_2)^2}\Bigg)^{1+i\gamma_{1s}}\numberthis\label{eq:LOcovamp}
\end{align*}
where we used that on the support of the delta function $u_s\.p_1=u_s\.p_2$, which implies that the conformal dimensions are the same, $\gamma_{1s}=\gamma_{2s}$. The fact that $u_s\.p_1=u_s\.p_2$ also verifies that the amplitude is symmetric under $p_1\leftrightarrow p_2$ despite first appearances in (\ref{eq:LOcovamp}). Amplitudes which exhibit this $\frac{1}{t^{1+i\gamma}}$, where $t=(p_{\text{in}}-p_{\text{out}})^2$, power-law behaviour have occurred in various instances across the physics literature when resumming certain subsets and subregions of perturbation theory in the eikonal limit \cite{Adamo:2021rfq, Kabat_1992}. However, let us emphasize two differences. The amplitude (\ref{eq:LOcovamp}) does not contain any implicit or explicit dimensional-regularization scale $\mu$ or cutoff scale $\Lambda_{\text{IR}}$. We note that the non-perturbative amplitudes for the scattering of massless particles on a self-dual Taub-NUT background \cite{Adamo:2023fbj} are also IR-finite and cutoff-independent. The second difference is that, when resumming subsets of perturbation theory, it is often not immediately apparent, or is at least highly non-trivial, to determine the NLO correction to that perturbation theory. Here, it is clear how to go to any order in the perturbation theory.\par
It is interesting to compare (\ref{eq:LOcovamp}) to the Coulomb-phase factor in the exponentiation theorem for the leading IR divergences (\ref{eq:Weinbergexponential}) in abelian gauge theories. Upon factoring out the tree-level amplitude $\frac{1}{t}$ from (\ref{eq:LOcovamp}), we see that the exponent in (\ref{eq:LOcovamp}) and the Coulomb-phase term $I$ in (\ref{eq:Weinbergexponential}) are both $\gamma_{1s}$. This similarity in the power-law behaviour suggests that the Coulomb-wavefunction perturbation theory at LO already contains the resummation of all the would-be leading infrared divergences. That is, the Coulomb-phase power-law behaviour of (\ref{eq:Weinbergexponentiation}) agrees with the LO amplitude's power-law behaviour, provided that we implement the following replacement in (\ref{eq:Weinbergexponentiation}):
\begin{gather}
    \Bigg(\frac{\lambda_{\text{IR}}}{\Lambda_{\text{IR}}}\Bigg)\rightarrow \frac{4\big(m^2-(p_1\.u_s)^2\big)}{(p_1-p_2)^2}=\frac{1}{\sin^2\frac{\theta}{2}},\label{replacement}
\end{gather}
 where $\theta$ is the scattering angle in the rest frame of the source particle. Throughout this paper we have restricted our attention to the Coulomb-phase divergence, ignoring the IR divergence associated with soft-photon production. It will be interesting to see whether the replacement (\ref{replacement}), applied to the $R$ term in (\ref{eq:Weinbergexponential}), also gives the correct leading power-law behaviour once the relevant IR-finite scattering amplitudes, which include soft-photon production, have been computed explicitly.
\subsection{Runge-Lenz symmetry of the LO amplitude}\label{sect:runge lenz}
\footnote{This section is outside the main focus of the article and can be skipped on a first reading.} As the LO amplitude (\ref{eq:LOcovamp}) is just a covariantization of the NRQM amplitude (\ref{treerestfframe}), which, as we reviewed in section \ref{sect:reviewrunge}, itself enjoys a Runge-Lenz symmetry, the LO amplitude also exhibits this symmetry.
We cannot expect this symmetry to extend beyond the LO amplitude as the $A^{\mu}A_{\mu}\phi$ term spoils this symmetry. \\
The covariantization of the Runge-Lenz symmetry follows from recalling that the Runge-Lenz symmetry does not affect the energy of the particles when working in the rest frame of the source particle. The covariantization of this is that the Runge-Lenz symmetry leaves the $u_s^{\mu}$ component of $p^{\mu}$ invariant. We therefore decompose the momenta as
\begin{gather}
    p^{\mu}=(p\.u_s)u_s^{\mu}+\sqrt{(p\.u_s)^2-m^2}\,\, \hat{p}^{\mu}_{\perp},
\end{gather}
where by construction $\hat{p}_{\perp}\. \hat{p}_{\perp}=-1$ and $\hat{p}_{\perp}\.u_s=0$. We then parameterize the subspace orthogonal to $u_s^{\mu}$ using polarization vectors
\begin{gather}
    \hat{p}_{\perp}^{\mu}=\sum_{i=1}^3\beta_i\e_i^{\mu}(u_s)\\
    \e_i\.\e_{j}=-\delta_{ij},\quad u_s\.\e_i=0,
\end{gather}
and because $\hat{p}_{\perp}\.\hat{p}_{\perp}=-1$ we have that $\vec{\beta}$ is a unit three-vector $|\vec{\beta}|^2=1$ which can be parameterized as
\begin{gather}
    \vec{\beta}(z)=\frac{1}{1+|z|^2}\Big(z+\bar{z},i(\bar{z}-z),1-|z|^2\Big),
\end{gather}
where $z$ is a function of $p^{\mu}$ and $u_s^{\mu}$ that depends on our choice of polarization vectors $\e_i^{\mu}$,
\begin{gather}
    z(p,u_s)=\frac{\hat{p}_{\perp}\.(\e_1+i \e_2)}{\hat{p}_{\perp}\.\e_3-1}.
\end{gather}
Altogether we have decomposed each momentum as
\begin{gather}
        p^{\mu}=(p\.u_s)u_s^{\mu}+\sqrt{(p\.u_s)^2-m^2}\,\, \sum_{i=1}^3\beta_i(z)\e_i^{\mu}(u_s)
\end{gather}
which allows us to write the $SO(1,3)$ Runge-Lenz transformation as
\begin{gather}
 \begin{pmatrix}
     a & b\\
     c & d
 \end{pmatrix}:   p^{\mu}\mapsto (p\.u_s)u_s^{\mu}+\sqrt{(p\.u_s)^2-m^2}\,\, \sum_{i=1}^3\beta_i\Big(\frac{az+b}{cz+d}\Big)\e_i^{\mu}(u_s),
\end{gather}
where $ad-bc=1$, and the matrix with entries $a,b,c,d$ here represents a group element of $SO(1,3)$. Under this transformation the LO amplitude (\ref{eq:LOcovamp}) transforms like a $2$-point correlation function in a 2D Euclidean CFT on a sphere, with the conformal dimension lying on the principal series, $\Delta=1+i\gamma$, and spin zero. We can then say that the cusp anomalous dimension $\gamma$ of abelian gauge theories is also the conformal dimension of the approximate/``tree-level'' Runge-Lenz conformal symmetry for the scattering of a spinless particle off a Coulomb background. It would be interesting to explore whether the Runge-Lenz symmetry of the Coulomb wavefunctions (\ref{eq:instate}, \ref{eq:outstate}) has any bearing on the ``infrared triangle'' of asymptotic symmetries, soft theorems, and memory effects for abelian gauge theories \cite{Strominger:2017zoo}.
\subsection{NLO amplitude}
The NLO amplitude reads
\begin{align*}
    &\mathcal{A}_1(p_1,p_2,u_s)\\
    &=\braket{\phi_{\text{out}}(p_2,u_s)|\hat{G}^{+}\hat{V}|\phi_{\text{in}}(p_1,u_s)}_{\text{SQED}}+\braket{\phi_{\text{out}}(p_2,u_s)|\hat{V}^{\dagger}\Big(\hat{G}^-\Big)^{\dagger}|\phi_{\text{in}}(p_1,u_s)}_{\text{SQED}}.\numberthis\label{NLO expression}
\end{align*}
Using that the first order perturbations satisfy $(\Box+m^2+2ie_1A^{\mu}\d_{\mu})_xG_{xy}V_y\phi_y=V_x\phi_x$, it is a straightforward calculation to check that the integrand here is again a divergenceless current when combining the two terms in (\ref{NLO expression}). We therefore have that (\ref{NLO expression}) is Lorentz invariant and independent of the Cauchy surface. Exploiting this freedom we choose to work in the rest frame of the source particle $u_s^{\mu}=\delta^{\mu}_0$ (which we will denote as $u_s^r$) and on an equal time $x^0$ Cauchy surface with normal vector $n^{\mu}=\delta^{\mu}_0$. Making these choices we obtain for the first term\footnote{On a first reading the reader may want to skip ahead to the intermediate result (\ref{exp}) as the following technical details obscure the simplicity of this result.}
\begin{align*}
&\braket{\phi_{\text{out}}(p_2,u_s)|\hat{G}^{+}\hat{V}|\phi_{\text{in}}(p_1,u_s)}_{\text{SQED}}=i\int\DD^3 x\int\DD^4 y\\
&\Bigg(\phi_{\text{out}}^{\star}(x,p_2)D_{x^0}G^{+}(x,y)V(y)\phi_{\text{in}}(y,p_1)
-\Big(D_{x^0}\phi_{\text{out}}(x,p_2)\Big)^{\star}G^{+}(x,y)V(y)\phi_{\text{in}}(y,p_1)\Bigg)\numberthis
\end{align*}
where we have explicitly indicated the meaning of all the terms and will now switch to a more compact and convenient notation. Choosing to express the retarded Green's function (\ref{eq:onshellgreens}) in terms of out-state wavefunctions, as will be useful momentarily, we find
\begin{align*}
    &\braket{\phi_{\text{out}}(p_2,u_s)|\hat{G}^{+}\hat{V}|\phi_{\text{in}}(p_1,u_s)}_{\text{SQED}}\\
    &=i\int\frac{\DD^3p}{(2\pi)^32E_p}\Big\langle \phi_{\text{out}}(p_2) \Big \vert \theta(x^0-y^0)\phi_{\text{out}}(E_p,\vec{p})\Big\rangle_{\text{SQED}}\braket{\phi_{\text{out}}(E_p,\vec{p})|\hat{V}|\phi_{\text{in}}(p_1)}_{\mathbb{R}^{1,3}}\\
    &-(E_p\rightarrow -E_p)\\
    &+
    \sum_{l=0}^{\infty}\sum_{n=l+1}^{\infty}\sum_{m=-l}^lE_n\Big\langle \phi_{\text{out}}(p_2) \Big \vert \theta(x^0-y^0)\phi_{nlm}\Big\rangle_{\text{SQED}}\braket{\phi_{nlm}|\hat{V}|\phi_{\text{in}}(p_1)}_{\mathbb{R}^{1,3}},\numberthis\label{onshellstates}
\end{align*}
where the $\mathbb{R}^{1,3}$ subscript on the inner products indicates integration over $\int\DD^4 y$. We have implemented a slight abuse of notation by placing the $\theta(x^0-y^0)$ time arguments inside the ket vectors; however, the notation is convenient for all of the other terms. The first two terms on the RHS of (\ref{onshellstates}) contain the continuum intermediate states, and the last term contains bound states at intermediate times. The SQED inner product contains a time derivative, which will act on $\theta(x^0-y^0)$ to give a delta function $\delta(x^0-y^0)$. In appendix \ref{app:vanishretardedgreens} we prove that the retarded Green's function vanishes at equal times
\begin{align*}
\delta(x^0-y^0) \Bigg\lbrace   &\sum_{l=0}^{\infty}\sum_{n=l+1}^{\infty}\sum_{m=-l}^lE_n\phi_{nlm}(x)\phi^{\star}_{nlm}(y)\\
&+\int\frac{\DD^3p}{(2\pi)^32E_p}\Bigg(\phi_{\text{in/out}}(p,x)\phi^{\star}_{\text{in/out}}(p,y)-(E_{p}\rightarrow -E_p)\Bigg)\Bigg\rbrace=0,\numberthis\label{usethisidentity}
\end{align*}
in which case the sum of the terms in (\ref{onshellstates}) where the time derivative acts on the theta function vanishes. Pulling the theta functions out of the SQED inner products in (\ref{onshellstates}), we find that only the positive-energy intermediate continuum state contributes to the amplitude, as the other two terms, the negative-energy and bound-state terms, have a vanishing SQED inner product with $\phi_{\text{out}}(p_2)$ due to energy conservation\footnote{We can now see why bound states only begin contributing at NNLO, as one requires two interaction vertices $V$: one to transition the continuum state into a bound state $\ket{B}$, and another to transition it back out, schematically $\braket{\phi_{\text{out}}|V|B}\braket{B|V|\phi_{\text{in}}}$.} We are left with
\begin{align*}
    &\braket{\phi_{\text{out}}(p_2,u^r_s)|\hat{G}^{+}\hat{V}|\phi_{\text{in}}(p_1,u^r_s)}_{\text{SQED}}\\
    &=i\int\frac{\DD^3p}{(2\pi)^32E_p}\theta(x^0-y^0)\braket{\phi_{\text{out}}(p_2) | \phi_{\text{out}}(p)}_{\text{SQED}}\braket{\phi_{\text{out}}(p)|\hat{V}|\phi_{\text{in}}(p_1)}_{\mathbb{R}^{1,3}}\numberthis\\
    &=i\alpha^2\int\DD^4 y\,\, \theta(x^0-y^0)\phi^{\star}_{\text{out}}(p_2,y)\frac{1}{|\vec{y}|^2}\phi_{\text{in}}(p_1,y)\numberthis\label{releasedelta}\\
    &=-\alpha^2\frac{1}{E_1-E_2+i\e}\int\DD^3\vec{y}\,\, \phi^{\star}_{\text{out}}(p_2,y)\frac{1}{|\vec{y}|^2}\phi_{\text{in}}(p_1,y)\at{y^0=x^0}\,\,\, ,\numberthis\label{theta}
\end{align*}
where $\alpha\defined \frac{e_1e_s}{4\pi}$. To go to line (\ref{releasedelta}) we first used the orthonormality of the Coulomb wavefunctions (\ref{orthonormal}), thus trivializing the $\DD^3\vec{p}$ integral. To go to line (\ref{theta}), we perform the $\int_{-\infty}^{x^0}\DD y^0$ integral, which is trivial in the rest frame of the source particle as all of the time dependence of the wavefunctions is in the $e^{-iEy^0}$ prefactors; see (\ref{restframeinstate}, \ref{restframeoutstate}). The $+i\e$ in the denominator of (\ref{theta}) is needed to provide the correct interpretation of the $-\infty$ lower bound of the $y^0$ integral and can be traced back to our use of the retarded Green's function for the in-state. Next we need to calculate the contribution from the corrections to the out-state, the second term in (\ref{NLO expression}), which instead uses the advanced Green's function. The intermediate steps are very similar, except that it is more convenient to express the Green's function in terms of in-states, and one instead encounters a $\theta(y^0-x^0)$. Performing these steps, we obtain the negative\footnote{There are three additional minus signs in this calculation: 1) The advanced Green's function (\ref{eq:onshellgreens}) has a minus sign, 2) We complex conjugate the Green's function to act on the bra vector, 3) The finite term comes from the lower $y^0$ integration bound as opposed to the upper bound previously.} of (\ref{theta}), except with a $-i\e$ instead (because this time we regulate an upper bound $\int^{\infty}_{x^0}\DD y^0$). Thus, when combining the two terms in (\ref{NLO expression}), we obtain
\begin{align}
     \mathcal{A}_1(p_1,p_2,u^r_s)&=-\alpha^2\Bigg(\frac{1}{E_1-E_2+i\e}-\frac{1}{E_1-E_2-i\e}\Bigg)\int\DD^3\vec{y}\,\, \phi^{\star}_{\text{out}}(p_2,y)\frac{1}{|\vec{y}|^2}\phi_{\text{in}}(p_1,y)\at{y^0=x^0}\label{oldfashioned}\\
     &=2\pi i\, \alpha^2\delta(E_1-E_2)\int\DD^3\vec{y}\,\, \phi^{\star}_{\text{out}}(p_2,y)\frac{1}{|\vec{y}|^2}\phi_{\text{in}}(p_1,y)\label{exp}
\end{align}
where it is no longer necessary to write $y^0=x^0$, as on the support of the energy-conserving delta function all $x^0$ dependence, which occurs as $e^{-ix^0(E_1-E_2)}$ in (\ref{exp}), drops out, as it must in order for the amplitude to be independent of the Cauchy surface. We note that (\ref{exp}) is much simpler than the steps leading up to it, and that (\ref{oldfashioned}) takes the form of a typical contribution to an amplitude in old-fashioned perturbation theory. This suggests that a Hamiltonian formulation of the perturbation theory should be simpler than the Green's function approach we have used here. We have avoided a Hamiltonian formulation here as the classical Hamiltonian of SQED requires additional machinery that we did not implement in order to not distract from the essential feature of this perturbation theory. However, for a fully quantum version of this perturbation theory, it would be better to use a Hamiltonian formulation. \par 
Heuristically, we expect (\ref{exp}) to be IR-finite as at large $|\vec{y}|$ distances the integrand takes the form of plane waves with a slight distortion factor (\ref{inexpansion}) scattering on a $\frac{1}{r^2}$ potential, and it is only the $\frac{1}{r}$ potential which gives rise to IR-divergences. We calculate (\ref{exp}) in appendix \ref{app:NLO integral}, and express the result in terms of its spherical harmonic expansion in order to compare with the exact amplitude in the next section. We find,
\begin{gather}
    \mathcal{A}_1(p_1,p_2,u^{r}_{s})=4\pi^2\alpha^2\frac{ \delta(E_1-E_2)}{|\vec{p}_1|}\sum_{l}\frac{\Gamma(1+l+i\gamma)}{\Gamma(1+l-i\gamma)}P_{l}(\cos\theta)\Big(i\pi+H_{l-i\gamma}-H_{l+i\gamma}\Big)\label{NLO spherical}
\end{gather}
where $H_n$ denotes the $n^{th}$ harmonic number. We perform the summation over angular momentum modes in appendix \ref{app:NLO summation}. There we are able to compute the first contribution to the amplitude in closed form
\begin{align*}
    \sum_{l=0}^{\infty}\frac{\Gamma(1+l+i\gamma)}{\Gamma(1+l-i\gamma)} &P_{l}(\hat{p}_1\.\hat{p}_2) =\\
    &\Bigg\lbrace\frac{1}{2}\frac{\Gamma\left(\frac{1}{2} + i \gamma\right) \,}{ \Gamma\left(\frac{1}{2} - i \gamma\right)}\csc\left(\frac{\theta}{2}\right)^{1 + 2i \gamma} {}_2F_1\left(\frac{1}{2}, -i \gamma, \frac{1}{2} - i \gamma, \sin^2\frac{\theta}{2}\right)\\ 
    &\quad  -\frac{1}{2i\gamma+1}\frac{\Gamma(i\gamma+1)}{\Gamma(-i\gamma)}\,\,{}_2F_1\left(\frac{1}{2}, 1 + i \gamma, \frac{3}{2} + i \gamma, \sin^2\frac{\theta}{2}\right)\Bigg\rbrace, \numberthis\label{NLOfirstexplict}
\end{align*}
whereas for the second set of terms in (\ref{NLO spherical}) we have only been able to reduce the sum to an integral 
\begin{align*}
 \sum_{l=0}^{\infty}\frac{\Gamma(1+l+i\gamma)}{\Gamma(1+l-i\gamma)}  P_{l}(\hat{p}_1\.\hat{p}_2) \Big(H_{l-i\gamma}-H_{l+i\gamma}\Big)=\frac{2^{ - 2i \gamma}}{\Gamma(-2i\gamma)}\int_0^{\infty}\DD \psi \frac{\psi\sinh(\psi)^{-1 - 2i \gamma}}{\sqrt{\sinh^2(\psi) + \sin^2\left(\frac{\theta}{2}\right)}}.\numberthis   \label{NLOsecondexplict}
\end{align*}
Both expressions (\ref{NLOfirstexplict}, \ref{NLOsecondexplict}) are IR-finite and do not contain any explicit or implicit IR cutoff. The integral (\ref{NLOsecondexplict}) can be readily evaluated numerically. In expressions (\ref{NLOfirstexplict}, \ref{NLOsecondexplict}), $\theta$ refers to the scattering angle in the rest frame of the source particle. The covariant expression for an arbitrary Lorentz frame is obtained upon making the replacement
\begin{gather}
    4\sin^2\frac{\theta}{2}\rightarrow \frac{(p_1-p_2)^2}{m^2-(u_s\.p_1)^2}\,.
\end{gather}
\subsection{Comparing to the exact result}\label{sect:exactresult}
The exact amplitude in the rest frame of the source particle is known in terms of its spherical-harmonic expansion (see appendix \ref{subapp:exactinout} for a derivation\footnote{The only derivation of this amplitude we were able to find is in Section 1.8 of \cite{pilkuhn2013relativistic}, where the discussion is framed in terms of extracting the amplitude from the asymptotics of the exact incoming wavefunction. In appendix \ref{subapp:exactinout} we calculate the amplitude instead in terms of the SQED inner product of the exact in/out states and find that the results agree with \cite{pilkuhn2013relativistic}.}) \cite{Adamo:2023cfp,Kol:2021jjc,PhysRevC.20.696,alma99741813502466}
\begin{align*}
   &\braket{\psi_{\text{out}}(p_2,u^r_s)|\psi_{\text{in}}(p_1,u^r_s)}_{\text{SQED}}\\
   &=\frac{4\pi^2\delta\big(E_1-E_2\big)}{|\vec{p}_1|}\sum_{l}(2l+1)\frac{\Gamma\Big(\sqrt{(l+\frac{1}{2})^2-\alpha^2}+\frac{1}{2}+i\gamma\Big)}{\Gamma\Big(\sqrt{(l+\frac{1}{2})^2-\alpha^2}+\frac{1}{2}-i\gamma\Big)}e^{i\pi\Big(l+\frac{1}{2}-\sqrt{(l+\frac{1}{2})^2-\alpha^2}\Big)}P_{l}(\cos\theta)\numberthis\label{eq:exactamp}
\end{align*}
Expanding this in powers of $\alpha$ gives
\begin{align*}
    &\braket{\psi_{\text{out}}(p_2,u^r_s)|\psi_{\text{in}}(p_1,u^r_s)}_{\text{SQED}}\\
    &=\frac{4\pi^2\delta\big(E_1-E_2\big)}{|\vec{p}_1|}\sum_{l}\frac{\Gamma(1+l+i\gamma)}{\Gamma(1+l-i\gamma)}P_{l}(\cos\theta)\Bigg[2l+1+\alpha^2\Big(i\pi+H_{l-i\gamma}-H_{l+i\gamma}\Big)\Bigg]+\O(\alpha^4),\numberthis\label{eq:exactexpansion}
\end{align*}
where $H_n$ denotes the $n^{th}$ harmonic number. We observe that the order $\alpha^{0}$ term matches our tree-level amplitude (\ref{tree:spherical}), and the order $\alpha^2$ term matches our NLO amplitude (\ref{NLO spherical}).\par
A comparison between the IR-divergent amplitude calculated up to third order in $\alpha$ using standard perturbation theory and the IR-finite exact expression (\ref{eq:exactamp}) at the same order can be found in \cite{kang1962higher}. They conclude that standard perturbation theory, up to the orders they compute, gives the exact result up to an overall IR-divergent phase. To be more precise, the one- and two-loop amplitudes for scattering on a $\frac{e^{-\lambda r}}{r}$ potential computed using standard perturbation theory contain factors $\log^{l}(\lambda/t)$, $l=1,2$, which diverge in the limit of a pure Coulomb potential, $\lambda\rightarrow 0$. The correct amplitude is obtained if one replaces $\lambda \rightarrow e^{\gamma_{E}}|\vec{p}|^2$, where $\gamma_{E}$ is Euler's constant, in the standard-perturbation-theory results.

\section{Conclusions}\label{sect:conc}
We have demonstrated that by using a basis of wavefunctions, namely in/out Coulomb wavefunctions, whose asymptotics appropriately match, in the precise sense of (\ref{keyassumption}), those of the exact in/out states, we are able to perturbatively compute \textit{the} IR-finite semiclassical scattering amplitude (\ref{eq:exactamp}).  We emphasize that the IR-finite semiclassical scattering amplitude is unambiguously defined, see Sect. \ref{sect:setup}, without reference to the perturbation theory one uses to construct it, and so is independent of our choice of Coulomb wavefunctions used for perturbatively computing it. We note that any choice of basis wavefunctions whose asymptotics appropriately match those of the exact in/out states should also yield an IR-finite perturbative framework for this semiclassical amplitude. In this paper we have only explored one such choice.  A natural choice, and one worth exploring, is the relativistic analog of the first term on the RHS of (\ref{inexpansion}), namely the distorted plane waves
\begin{align*}
    &\phi_{\text{in}}(p,x,u_s)\\
    &=\exp\Bigg(ip\. x+i\gamma\Big(\sqrt{(p\.u_s)^2-p^2}\sqrt{(u_s\.x)^2-x^2}-(p\.u_s)(u_s\.x) +(p\.x)\Big)\Bigg),\numberthis\label{maybethis}
\end{align*}
which would be a relativistic generalization of the work by Mulherin and Zinnes \cite{mulherin1970coulomb, barrachina1989scattering,Kadyrov_2005}. One can characterize the difference between the FK approach and Coulomb wavefunction perturbation theory as the former modifies the evolution operators in the key equation (\ref{keyassumption}), whereas the latter modifies the states.\par
Let us briefly outline which aspects of the perturbation theory we expect to be relevant in more realistic settings, such as QED, where the fields are quantized and the gauge field is dynamical rather than a fixed background. Incorporating the spin of the electron is straightforward, as the solution to the ``semi-free'' Dirac equation is well known \cite{Berestetskii1982118}. We expect the early-time dynamics of the exact in-states in QED to differ from that of free-particle states in two respects: first, charged particles move in the fixed Coulomb backgrounds of all the other charged particles; and second, there is a cloud of soft photons surrounding pairs of charged particles. If only the former difference occurred, then the Coulomb wavefunction perturbation theory would carry over straightforwardly to QED. However, the extent to which these two processes interfere makes it unclear whether Coulomb wavefunctions \textit{per se} are the appropriate wavefunctions to use in QED; this remains a topic for future research.
\par
In summary, standard perturbation theory views a scattering event as free particles meeting at vertices and transitioning to other free particles, while Coulomb wavefunction perturbation theory views a scattering process as Coulomb states meeting at vertices and transitioning to other Coulomb states.
\section*{Acknowledgments}
We thank Tim Adamo, Hofie Hannesdottir, and Anton Ilderton for providing valuable feedback on a draft version of the manuscript. It is a pleasure to acknowledge insightful conversations with Andrew McLeod, Marcus Spradlin, and Akshay Srikant. We thank CERN for its hospitality while this work was partially completed. This work is supported by the enhanced research expenses \texttt{RF\textbackslash ERE\textbackslash 221030} associated with the Royal Society grant \texttt{URF\textbackslash R1\textbackslash 221233}. OpenAI Codex was used solely to assist with proofreading and checks of LaTeX formatting, bibliography, equation formatting and internal consistency, and cross-references. All suggested changes were reviewed by the author. It was not used to generate the scientific content.
\appendix
\section{The exact amplitude}\label{app:exactamp}
In appendix \ref{subapp:exactinout} we compute the exact in/out wavefunctions for a KG particle scattering on a Coulomb background, as well as the exact non-perturbative amplitude (\ref{exactampinapp}). In appendix \ref{subapp:verifyasymp} we verify that the Coulomb wavefunctions (\ref{eq:instate}, \ref{eq:outstate}) satisfy the key requirement (\ref{keyassumption}) that their asymptotics appropriately match those of the exact in/out states.
\subsection{The exact in/out states and exact scattering amplitude}\label{subapp:exactinout}
It is straightforward to find the spherically symmetric solutions to the full SQED equation of motion in the rest frame of the source particle
\begin{gather}
\Bigg(\Box +m^2+2i\frac{\alpha}{|\vec{x}|}\d_t-\frac{\alpha^2}{|\vec{x}|^2}\Bigg)e^{-iE_kt}\psi^{|\vec{k}|}_{lm}\Big(\vec{x},u^{r}_s\Big)=0\\
    \psi^{|\vec{k}|}_{lm}\Big(\vec{x},u^{r}_s\Big)=Y_{lm}(\hat{x})R_l\big(|\vec{k}|,|\vec{x}|\big) \label{psiexact}
\end{gather}
as the radial equation is solved by\footnote{This solution can be derived from the spherically symmetric solution in the absence of the $\frac{\alpha^2}{r^2}$ term, as this term merely shifts the angular momentum variable $l$ in the radial equation.}
\begin{align*}
 &R_l\big(|\vec{k}|,|\vec{x}|\big)   \\
 &=\frac{|\vec{k}|}{\sqrt{\pi E}}e^{-\frac{\pi\gamma}{2}}\frac{\Gamma(\frac{1}{2}+s+i\gamma)}{\Gamma(1+2s)}(2i|\vec{k}||\vec{x}|)^{s-\frac{1}{2}}e^{-i|\vec{k}||\vec{x}|}{}_1F_1\Bigg(\frac{1}{2}+s-i\gamma,1+2s,2i|\vec{k}||\vec{x}|\Bigg)\numberthis\label{radialexact}
 \end{align*}
 \begin{gather}
 s\defined\sqrt{\Big(l+\frac{1}{2}\Big)^2-\alpha^2}
\end{gather}
and we chose the normalization here so that these functions are orthonormal with respect to the SQED ``inner product'' \cite{1964JMP.....5..591H}
\begin{gather}
    \Big\langle \psi^{|\vec{k}'|}_{l'm'}\big(u_s^r\big)\, \Big|\,  \psi^{|\vec{k}|}_{lm}\big(u_s^r\big)\Big\rangle_{\text{SQED}}=\delta_{ll'}\delta_{mm'}\delta\Big(|\vec{k}|'-|\vec{k}|\Big).\label{orthoexact}
\end{gather}
We now turn to constructing the exact in and out states,
\begin{gather}
  \psi_{\text{in/out}}(k^{\mu},x^{\mu},u_s^r)=e^{-iE_kt}\sum_{l,m}C_{\text{in/out}}(\vec{k},l,m)\psi^{|\vec{k}|}_{lm}\Big(\vec{x},u^{r}_s\Big) \label{expansionsphericalexact}
\end{gather}
where the coefficients $C_{\text{in/out}}$ are determined by the conditions that $\psi_{\text{in/out}}$ appear as much as possible like a plane wave at early/late times. To be more precise, the large $|\vec{x}|$ asymptotics of a plane wave is given by
\begin{gather}
    \lim_{|\vec{x}|\rightarrow \infty}e^{i\vec{k}\.\vec{x}}= \frac{1}{|\vec{k}||\vec{x}|}\sum_{l=0}^{\infty}i^{l}(2l+1)P_{l}(\hat{k}\.\hat{x})\sin\Big(|\vec{k}||\vec{x}|-\frac{l\pi}{2}\Big)\label{plane wave}
\end{gather}
which is composed of incoming waves $e^{-i|\vec{k}||\vec{x}|+il\pi/2}$ and outgoing waves $e^{i|\vec{k}||\vec{x}|-il\pi/2}$. Similarly, the large $|\vec{x}|$ asymptotics of the radial wavefunction (\ref{radialexact}) is composed of an incoming $e^{-i|\vec{k}||\vec{x}|}$ wave and an outgoing $e^{i|\vec{k}||\vec{x}|}$ wave, where the exact form of these two terms can be obtained by performing an asymptotic expansion of the radial wavefunction (\ref{radialexact}) using the asymptotics of the confluent hypergeometric function (\ref{asymp}). We choose $C_{\text{in}}$ so that the incoming waves in $\psi_{\text{in}}$ agree as much as possible with the incoming waves of the plane wave (\ref{plane wave}), and we choose $C_{\text{out}}$ so that the outgoing waves in $\psi_{\text{out}}$ agree as much as possible with the outgoing waves in (\ref{plane wave}). Doing so, we find
\begin{gather}
      C_{\text{in}}(\vec{k},l,m)=4\pi i\frac{\sqrt{\pi E}}{|\vec{k}|}(-1)^le^{-i\pi s}Y^{\star}_{lm}(\hat{k})\label{in coeff}\\
     C_{\text{out}}(\vec{k},l,m)=4\pi \frac{\sqrt{\pi E}}{|\vec{k}|}\frac{\Gamma(s+\frac{1}{2}-i\gamma)}{\Gamma(s+\frac{1}{2}+i\gamma)}Y^{\star}_{lm}(\hat{k})\label{out coeff}
\end{gather}
Thus combining (\ref{psiexact}, \ref{radialexact}, \ref{expansionsphericalexact}, \ref{in coeff}, \ref{out coeff}) we have explicitly constructed the exact in/out states in the rest frame of the source particle. We can now easily calculate the exact scattering amplitude using the orthonormality of the basis wavefunctions (\ref{orthoexact}),
\begin{align*}
    &\braket{\psi_{\text{out}}(p_2^{\mu},u^r_s)|\psi_{\text{in}}(p_1^{\mu},u^r_s)}_{\text{SQED}}\\
   &=\frac{4\pi^2\delta\big(E_1-E_2\big)}{|\vec{p}_1|}\sum_{l}(2l+1)\frac{\Gamma\Big(\sqrt{(l+\frac{1}{2})^2-\alpha^2}+\frac{1}{2}+i\gamma\Big)}{\Gamma\Big(\sqrt{(l+\frac{1}{2})^2-\alpha^2}+\frac{1}{2}-i\gamma\Big)}e^{i\pi\Big(l+\frac{1}{2}-\sqrt{(l+\frac{1}{2})^2-\alpha^2}\Big)}P_{l}(\hat{p}_1\.\hat{p}_2)\numberthis\label{exactampinapp}
\end{align*}
where one must complex conjugate the $C_{\text{out}}$ coefficients when taking this inner product, and we also used that
\begin{gather}
    P_{l}(\hat{k}'\.\hat{k})=\frac{4\pi}{2l+1}\sum_{m=-l}^lY^{\star}_{lm}(\hat{k}')Y_{lm}(\hat{k}).
\end{gather}
We remind the reader that all expressions are understood to be in the rest frame of the source particle.
\subsection{Verifying that the asymptotics agree}\label{subapp:verifyasymp}
With the exact in/out states in hand, we can explicitly verify that the key assumption (\ref{keyassumption}) is satisfied. 
Using the explicit form of the exact in/out states given in the previous appendix \ref{subapp:exactinout}, as well as using the asymptotics of the confluent hypergeometric function (\ref{asymp}), we have that, in the rest frame of the source particle (which we denote as $u_s^r$), the asymptotics of the exact in/out states read,
\begin{align}
    \lim_{|\vec{x}|\rightarrow \infty}\psi_{\text{in}}(k^{\mu},x^{\mu},u_s^r)&= i e^{-i|\vec{k}||\vec{x}|}\Big(2|\vec{k}||\vec{x}|\Big)^{-1+i\gamma}\sum_{l=0}^{\infty}(-1)^l(2l+1)P_{l}(\hat{k}\.\hat{x}) +\Big(\substack{\text{Outgoing} \\ \text{scattered waves}}\Big)\label{incomingexact}\\
        \lim_{|\vec{x}|\rightarrow \infty}\psi_{\text{out}}(k^{\mu},x^{\mu},u_s^r)&=(-i) e^{i|\vec{k}||\vec{x}|}\Big(2|\vec{k}||\vec{x}|\Big)^{-1-i\gamma}\sum_{l=0}^{\infty}(2l+1)P_{l}(\hat{k}\.\hat{x}) +\Big(\substack{\text{Incoming} \\ \text{scattered waves}}\Big),
\end{align}
where the ``outgoing/incoming scattered waves'' terms are not explicitly given here, as they are not needed for the present purpose of verifying that the key assumption (\ref{keyassumption}) is satisfied. The key assumption only requires that the early-time asymptotics of the exact in-wavefunction match the early-time asymptotics of the input in-wavefunction (\ref{eq:instate}). A stationary-phase analysis at asymptotically early times will pick out the incoming waves in the incoming states (\ref{incomingexact}, \ref{incomingperturb}); hence, it is only these incoming waves that need to agree for the in-states. The argument for the out-states is similar. Note that the incoming waves in the in-state (\ref{incomingexact}) are independent of the $\alpha^2=(\frac{e_1e_2}{4\pi})^2$ coupling, which is a proxy for the $A^2\phi^2$ quartic vertex. This verifies our expectation that the incoming wave of the in-state can be chosen to be independent of the details of the $\frac{\alpha^2}{|\vec{x}|^2}$ term in the potential. We can compute the large $|\vec{x}|$ asymptotics of the Coulomb wavefunctions (\ref{eq:instate}, \ref{eq:outstate}) using the Legendre-polynomial expansion of these wavefunctions given in (\ref{eq:partialwavein}, \ref{eq:sphericaldecomp}) and the asymptotics of the confluent hypergeometric function (\ref{asymp}) to find
\begin{align}
   \lim_{|\vec{x}|\rightarrow \infty} \phi_{\text{in}}(k^{\mu},x^{\mu},u_s^r)&=i(2|\vec{k}||\vec{x}|)^{-1+i\gamma}e^{-i|\vec{k}||\vec{x}|}\sum_l(-1)^l(2l+1)P_{l}(\hat{x}\.\hat{k})+\Big(\substack{\text{Outgoing} \\ \text{scattered waves}}\Big)\label{incomingperturb}\\
      \lim_{|\vec{x}|\rightarrow \infty}  \phi_{\text{out}}(k^{\mu},x^{\mu},u_s^r)&=(-i)(2|\vec{k}||\vec{x}|)^{-1-i\gamma}e^{i|\vec{k}||\vec{x}|}\sum_l(2l+1)P_{l}(\hat{x}\.\hat{k})+\Big(\substack{\text{Incoming} \\ \text{scattered waves}}\Big)
\end{align}
By noting that (\ref{incomingexact}) $=$ (\ref{incomingperturb}) we see that the incoming waves of the Coulomb wavefunctions and the incoming waves of the exact in-wavefunctions agree, thus verifying that the key condition (\ref{keyassumption}) is satisfied for the in-states. Similarly for the out-states.
\section{Integrals and summations}\label{app:LO Int}
In this appendix we will compute integrals that involve products of confluent hypergeometric functions. There have been several studies of this class of integrals \cite{nordsieck1954reduction,saad2003integrals,colavecchia1997hypergeometric,gravielle1992some}, although we will only make use of appendix $f$ of \cite{Landau:1991wop} for the evaluation of one of the integrals.
\subsection{Orthonormality of the Coulomb-wavefunctions}\label{subapp:ortho}
In this appendix we demonstrate that
\begin{gather}
    \braket{\phi_{\text{in}}(p_1)|\phi_{\text{in}}(p_2)}_{\text{SQED}}=\braket{\phi_{\text{out}}(p_1)|\phi_{\text{out}}(p_2)}_{\text{SQED}}=(2\pi)^32 E_1\delta^3(\vec{p}_1-\vec{p}_2)  \label{apporthonormal}
\end{gather}
where the Coulomb wavefunctions $\phi_{\text{in/out}}$ are those given in (\ref{eq:instate}, \ref{eq:outstate}). The intermediate results obtained in this subsection will allow us to evaluate the LO amplitude integral in the next subsections in short succession. We exploit the Lorentz invariance and Cauchy-slice independence of (\ref{apporthonormal}) to work in the rest frame of the source particle $u_{s}^r$ on an equal-time Cauchy slice. The wavefunctions in the rest frame of the source particle are given in (\ref{restframeinstate}, \ref{restframeoutstate}). We expand these rest-frame wavefunctions in terms of spherical harmonics\footnote{We note that Nordsieck's method \cite{nordsieck1954reduction} could be applied to evaluate this integral without going to the spherical-harmonic basis.}
\begin{align}
    \phi_{\text{in}}(x,p,u_s^r)&=e^{-\frac{\pi\gamma}{2}}e^{-iEt}\sum_{l=0}^{\infty}\frac{\Gamma(1+l+i\gamma)}{(2l)!}(2i|\vec{p}|r)^le^{-i|\vec{p}|r}{}_1F_1\Big(1+l-i\gamma,2l+2,2i|\vec{p}|r\Big)P_{l}(\hat{x}\.\hat{p})\label{eq:partialwavein}\\
        \phi_{\text{out}}(x,p,u_s^r)&=e^{-\frac{\pi\gamma}{2}}e^{-iEt}\sum_{l=0}^{\infty}\frac{\Gamma(1+l-i\gamma)}{(2l)!}(2i|\vec{p}|r)^le^{-i|\vec{p}|r}{}_1F_1\Big(1+l-i\gamma,2l+2,2i|\vec{p}|r\Big)P_{l}(\hat{x}\.\hat{p})\label{eq:sphericaldecomp}
\end{align}
where $\hat{x}\.\hat{p}=\frac{\vec{x}\.\vec{p}}{|\vec{x}||\vec{p}|}$. Using these in (\ref{apporthonormal}), the resultant angular integral can then be straightforwardly computed using
\begin{gather}
    \int\DD^2 \Omega\,\, P_{l}(\hat{k}_1\.\hat{x})P_{l'}(\hat{k}_2\.\hat{x})=\delta_{ll'}\frac{4\pi}{2l+1}P_{l}(\hat{k}_1\.\hat{k}_2).
\end{gather}
There are then two pieces to the remaining integral
\begin{align*}
  &\braket{\phi_{\text{in}}(p_2,u_s^r)|\phi_{\text{in}}(p_1,u_s^r)}_{\text{SQED}}=4 \pi  e^{-\frac{\pi}{2}(\gamma_1+\gamma_2)}e^{-it(E_1-E_2)}\\
  &\sum_{l=0}^{\infty} \frac{\Gamma(1+l+i\gamma_{1})\Gamma(1+l-i\gamma_2)}{(2l+1)[(2l)!]^2} (4|\vec{p}_1||\vec{p}_2|)^l P_l(\hat{p}_1\.\hat{p}_2)\Bigg(\text{T}_1\Big(|\vec{p}_1|,|\vec{p}_2|,l\Big)+\text{T}_2\Big(|\vec{p}_1|,|\vec{p}_2|,l\Big)\Bigg)\numberthis\label{startortho}
\end{align*}
where the two terms are
\begin{align*}
\text{T}_1(|\vec{p}_1|,|\vec{p}_2|,l)&=(E_1+E_2)\int_0^{\infty}\DD r\,\, r^{2+2l}e^{-ir(|\vec{p}_1|-|\vec{p}_2|-i\e_{+})}\\
&{}_1F_1\Big(1+l-i\gamma_1,2l+2,2i|\vec{p}_1|r\Big){}_1F_1\Big(1+l+i\gamma_2,2l+2,-2i|\vec{p}_2|r\Big)\numberthis\label{T1def}\\
\text{T}_2(|\vec{p}_1|,|\vec{p}_2|,l)&=-2\frac{e_1e_2}{4\pi}\int_0^{\infty}\DD r\,\, r^{1+2l}e^{-ir(|\vec{p}_1|-|\vec{p}_2|-i\e_{+})}\\
&{}_1F_1\Big(1+l-i\gamma_1,2l+2,2i|\vec{p}_1|r\Big){}_1F_1\Big(1+l+i\gamma_2,2l+2,-2i|\vec{p}_2|r\Big)\numberthis\label{T2 int}
\end{align*}
where we have added a positive infinitesimal $\e_{+}$ to regulate the above integrals, which we will take to zero at the end of the calculation\footnote{We emphasize that the difference between this infinitesimal and a traditional IR cutoff $\Lambda_{\text{IR}}$, or a dimensional-regularization parameter, is that we will be able to take $\e_{+}$ to zero at the end of the calculation without encountering any divergence.} Note that $\gamma_1$ and $\gamma_2$ are not necessarily equal because $\gamma$ is energy-dependent. The second of these integrals, (\ref{T2 int}), is evaluated using Eqs. (f.9) and (f.10) of \cite{Landau:1991wop}.
\begin{align*}
    \int_0^{\infty}\DD z\,\,& e^{-\lambda z}z^{\gamma-1}{}_1F_1(\alpha,\gamma, kz){}_1F_1(\alpha',\gamma,k'z)\\
    &=\Gamma(\gamma)\lambda^{\alpha+\alpha'-\gamma}(\lambda-k)^{-\alpha}(\lambda-k')^{-\alpha'}{}_2F_1\Bigg(\alpha,\alpha',\gamma,\frac{kk'}{(\lambda-k)(\lambda-k')}\Bigg)\numberthis\label{eq:Landauint}
\end{align*}
which gives
\begin{align*}
  \text{T}_2(|\vec{p}_1|,|\vec{p}_2|,l)=&-2\frac{e_1e_2}{4\pi}e^{\pi \gamma_1} \Gamma(2 + 2 l)\Big( |\vec{p}_1| - |\vec{p}_2| \Big)^{-i (\gamma_1 - \gamma_2)} \Big( |\vec{p}_1| + |\vec{p}_2| \Big)^{-2 - 2l + i (\gamma_1 -  \gamma_2)}  \\
  &{}_2F_1\Big(1 + l - i \gamma_1, 1 + l + i \gamma_2, 2l+2, \frac{4 |\vec{p}_1| |\vec{p}_2|}{( |\vec{p}_1| + |\vec{p}_2| )^2} \Big)\numberthis
\end{align*}
where we took $\e_{+}\rightarrow 0$ and did not encounter any distributional terms in taking this limit.
The $\text{T}_1$ integral (\ref{T1def}) can then be obtained by differentiating (\ref{eq:Landauint}) with respect to $\lambda$. Doing so, we obtain two terms for $T_1$, a distributional piece $R_1$ and a non-distributional piece $R_2$,
\begin{align*}
    \text{T}_1(|\vec{p}_1|,|\vec{p}_2|,l)&=(E_1+E_2)(R_1+R_2)\numberthis\label{T1istwoterms}
\end{align*}
where 
\begin{align*}
 R_2(|\vec{p}_1|,|\vec{p}_2|,l)&= 2 e^{\pi \gamma_1} \Gamma(2 + 2 l)\Big( |\vec{p}_1| - |\vec{p}_2| \Big)^{-1 - i (\gamma_1 - \gamma_2)} \Big( |\vec{p}_1| + |\vec{p}_2| \Big)^{-3 - 2l + i (\gamma_1 -  \gamma_2)} \\
  &\times \Big( |\vec{p}_1| \gamma_1 - |\vec{p}_2| \gamma_2 \Big) {}_2F_1\Big(1 + l - i \gamma_1, 1 + l + i \gamma_2, 2(1+l), \frac{4 |\vec{p}_1| |\vec{p}_2|}{( |\vec{p}_1| + |\vec{p}_2| )^2} \Big)\numberthis.\label{diesinnorqm}
\end{align*}
\footnote{Note that in the non-relativistic limit where $\gamma=\alpha\frac{E}{|\vec{p}|}\rightarrow \alpha\frac{m}{|\vec{p}|} $ we have that (\ref{diesinnorqm})$=0$ due to $|\vec{p}_1| \gamma_1 - |\vec{p}_2| \gamma_2\rightarrow 0$. This demonstrates that the NRQM Coulomb wavefunctions are orthogonal with respect to the $\int\DD^3 x$ inner product.} Some straightforward algebra using the on-shell condition $p^2=m^2$ then demonstrates the cancellation,
\begin{gather}
    (E_1+E_2)R_2+T_2=0.\label{cancellation}
\end{gather}
We are thus left with only the distributional term

\begin{align*}
    R_1&=2e^{\frac{\pi}{2} (\gamma_1 + \gamma_2)} \Gamma(2 + 2 l)\lim_{\e_{+}\rightarrow 0}  \e_{+}  \Big( i (|\vec{p}_1| - |\vec{p}_2|) + \e_{+} \Big)^{-i (\gamma_1 - \gamma_2)} \Big( |\vec{p}_1| + |\vec{p}_2| \Big)^{-4 - 2l + i (\gamma_1 -  \gamma_2)} \\
  &\times \Big( 1 + l - i \gamma_1 \Big)\, {}_2F_1\Big(2 + l - i \gamma_1, 1 + l + i \gamma_2, 2(1+l), \frac{4 |\vec{p}_1| |\vec{p}_2|}{( |\vec{p}_1| + |\vec{p}_2| )^2+\e_{+}^2} \Big)\numberthis
\end{align*}
In order to get a non-zero result in the limit $\e_{+}\rightarrow 0$, we need the last argument of ${}_2F_1$ to approach $1$, which occurs at $|\vec{p}_1|=|\vec{p}_2|$. We can then use Eq. (15.4.23) of \url{https://dlmf.nist.gov/15.4#ii}
\begin{gather}
    \lim_{z\to 1^{-}}\frac{F\left(a,b;c;z\right)}{(1-z)^{c-a-b}}=\frac{\Gamma\left(c%
\right)\Gamma\left(a+b-c\right)}{\Gamma\left(a\right)\Gamma\left(b\right)},\qquad \text{Re}[c-a-b]<0
\end{gather}
and the relation
\begin{equation}
    \lim_{\e\rightarrow 0}\frac{\e}{x^2+\e^2}=\pi\delta(x)
\end{equation}
to find
\begin{align}
    R_1&=4\pi\delta\Big(|\vec{p}_1|-|\vec{p}_2|\Big)\frac{2^{-3 - 2 l} e^{\pi \gamma_1} |\vec{p}_1|^{-2l-2} \Gamma(2 + 2 l)^2}{\Gamma(1 + l - i \gamma_1) \Gamma(1 + l + i \gamma_1)}
\end{align}
where we used that on the support of the delta function $\gamma_1=\gamma_2$. Returning to (\ref{startortho}) we have
\begin{align}
      \braket{\phi_{\text{in}}(p_2,u_s^r)|\phi_{\text{in}}(p_1,u_s^r)}_{\text{SQED}}&=\delta(|\vec{p}_1|-|\vec{p}_2|)4\pi^2 \frac{E_1}{|\vec{p}_1|^2}\sum_{l}(2l+1)P_{l}(\hat{p}_1\.\hat{p}_2)\label{eq:tomultiply}\\
      &=16\pi^2\frac{E_1}{|\vec{p}_1|^2}\delta(|\vec{p}_1|-|\vec{p}_2|) \delta(1-\hat{p}_1\.\hat{p}_2)\\
      &=(2\pi)^3 2E_1\delta^3(\vec{p}_1-\vec{p}_2) 
\end{align}
We note that the result is already Lorentz invariant so there is no need to ``covariantize'' for $u_s^{\mu}$ not in the rest frame.

\subsection{LO amplitude integral}\label{app:LOamp}
In this section we compute the integral for the LO amplitude
\begin{align}
  \mathcal{A}_0(p_2,p_1;u_s)&=\braket{\phi_{\text{out}}(p_2,u_s)|\phi_{\text{in}}(p_1,u_s)}_{\text{SQED}}\\
 &=i\int_{\Sigma}\DD^3\Sigma_{\mu}\, \Big[\phi_{\text{out}}^{\star}(x,p_2)D^{\mu}\phi_{\text{in}}(x,p_1)-\phi_{\text{in}}(x,p_1)\Big(D^{\mu}\phi_{\text{out}}(x,p_2)\Big)^{\star}\Big]
\end{align}
where the Coulomb wavefunctions $\phi_{\text{in/out}}$ are those given in (\ref{eq:instate}, \ref{eq:outstate}). We will compute this integral in terms of its spherical-harmonic expansion in order to compare it with the known exact result in section \ref{sect:exactresult}. Using the same approach of expanding in terms of spherical harmonics in the rest frame of the source particle as in appendix \ref{subapp:ortho}, the amplitude reads
\begin{align*}
  &\braket{\phi_{\text{out}}(p_2,u_s^r)|\phi_{\text{in}}(p_1,u_s^r)}_{\text{SQED}}=4 \pi  e^{-\frac{\pi}{2}(\gamma_1+\gamma_2)}e^{-it(E_1-E_2)}\\
  &\sum_{l=0}^{\infty} \frac{\Gamma(1+l+i\gamma_{1})\Gamma(1+l+i\gamma_2)}{(2l+1)[(2l)!]^2} (4|\vec{p}_1||\vec{p}_2|)^l P_l(\hat{p}_1\.\hat{p}_2)\Bigg(\text{T}_1\Big(|\vec{p}_1|,|\vec{p}_2|,l\Big)+\text{T}_2\Big(|\vec{p}_1|,|\vec{p}_2|,l\Big)\Bigg)\numberthis\label{startLOamp}
\end{align*}
where $\text{T}_{1,2}$ are given in (\ref{T1def}, \ref{T2 int}). The only difference between the orthonormalization integral (\ref{startortho}) and (\ref{startLOamp}) is the sign of $\gamma_{2}$ in the second Gamma-function factor in the numerator, i.e. $\Gamma(1+l\pm i\gamma_{2})$. Multiplying (\ref{eq:tomultiply}) by $\frac{\Gamma(1+l+i\gamma_2)}{\Gamma(1+l-i\gamma_2)}$ in the appropriate fashion then gives us the result
\begin{align}
      \braket{\phi_{\text{out}}(p_2,u_s^r)|\phi_{\text{in}}(p_1,u_s^r)}_{\text{SQED}}&=\delta(|\vec{p}_1|-|\vec{p}_2|)4\pi^2 \frac{E_1}{|\vec{p}_1|^2}\sum_{l}(2l+1)\frac{\Gamma(1+l+i\gamma_1)}{\Gamma(1+l-i\gamma_1)}P_{l}(\hat{p}_1\.\hat{p}_2)\label{LOsumintermed}
\end{align}
where we used $\gamma_1=\gamma_2$ on the support of the delta function.
\subsection{NRQM amplitude integral}\label{app:NRQM amp integral}
In this appendix, we compute the NRQM scattering amplitude by evaluating the $\mathrm{d}^3x$ inner product between the in-state (\ref{NRQMinstate}) and the out-state (\ref{NRQMoutstate}). Specifically, we show that
\begin{align}
    A^{\text{NRQM}}(\vec{p}_1,\vec{p}_2) &= \int \mathrm{d}^3 x\, \phi^{\star}_{\text{out}}(\vec{x},\vec{p}_2)\phi_{\text{in}}(\vec{x},\vec{p}_1) \label{inoutNRQMamp}\\
    &=-\frac{i\pi}{2}\frac{e_1e_2}{|\vec{p}|^2}\delta(E_1-E_2)\frac{\Gamma(1+i\gamma_1)}{\Gamma(1-i\gamma_1)}\frac{1}{\sin^2(\theta/2)}e^{-i\gamma\log\sin^2(\theta/2)},\quad \theta \neq 0.\label{ampNRQMint}
\end{align}
The NRQM in/out states $\phi_{\text{in/out}}$ in (\ref{inoutNRQMamp}) differ from the relativistic Coulomb wavefunctions (\ref{restframeinstate}, \ref{restframeoutstate}) only in their energy dependence, where $E = \frac{|\vec{p}|^2}{2m}$ is replaced by $\sqrt{|\vec{p}|^2 + m^2}$, and in their conformal dimension, which shifts from $\gamma = \frac{e_1 e_2}{4\pi} \frac{m}{|\vec{p}|}$ to $\frac{e_1 e_2}{4\pi} \frac{\sqrt{|\vec{p}|^2 + m^2}}{|\vec{p}|}$. It is convenient to express the NRQM in/out wavefunctions in the partial-wave basis as
\begin{align}
    \phi_{\t{in/out}}(\vec{x},\vec{p})&=\frac{1}{2|\vec{p}|}\sum_{l=0}^{\infty}i^l(2l+1)e^{\pm i\delta_l}R_l(|\vec{p}|,r)P_{l}(\hat{x}\.\hat{p})\label{NRQM partial wave}\\
    R_l(|\vec{p}|,r)&=2|\vec{p}|e^{-\frac{\pi\gamma}{2}}\frac{|\Gamma(1+l+i\gamma)|}{(2l+1)!}(2|\vec{p}|r)^le^{i|\vec{p}|r}{}_1F_1(1+l+i\gamma,2l+2,-2i|\vec{p}|r)\label{NRQM radial}\\
    e^{2i\delta_l}&=\frac{\Gamma(1+l+i\gamma)}{\Gamma(1+l-i\gamma)},
\end{align}
where the phase factors $e^{+i\delta_l}$ and $e^{-i\delta_l}$ correspond to the in-state and out-state, respectively, in (\ref{NRQM partial wave}). For a proof of (\ref{NRQM partial wave}) see, for example, Section 10, Chapter 11 of \cite{messiah1999quantum}. The $\mathrm{d}^3x$ inner product between the in-state and out-state can then be computed directly using
\begin{gather}
    \int \DD^2\Omega P_l(\hat{p}_1\.\hat{x})P_{l'}(\hat{p}_2\.\hat{x})=\delta_{ll'}\frac{4\pi}{2l+1}P_{l}(\hat{p}_1\.\hat{p}_2),\label{angular sum of legendres}
\end{gather}
to evaluate the angular integral. We can then use the orthonormality of the radial wavefunctions (\ref{NRQM radial}) (see Section 136 of \cite{Landau:1991wop}) to compute the radial integral
\begin{gather}
    \int_0^{\infty}\DD r\, r^2 R_l(|\vec{p}_1|,r)R_l(|\vec{p}_2|,r)=2\pi \delta(|\vec{p}_1|-|\vec{p}_2|).
\end{gather}
Altogether one obtains
\begin{align}
     \int \mathrm{d}^3 x\, \phi^{\star}_{\text{out}}(\vec{x},\vec{p}_2)\phi_{\text{in}}(\vec{x},\vec{p}_1)&=\delta(|\vec{p}_1|-|\vec{p}_2|) \frac{2\pi^2}{|\vec{p}_1|^2}\sum_{l}(2l+1)\frac{\Gamma(1+l+i\gamma)}{\Gamma(1+l-i\gamma)}P_{l}(\hat{p}_1\.\hat{p}_2)\label{almost done}
\end{align}
The sum over partial waves can be evaluated using \cite{mott1932polarisation, Lin_2000}:
\begin{gather}
    \sum_{l=0}^{\infty}(2l+1)\frac{\Gamma(1+l+i\gamma)}{\Gamma(1+l-i\gamma)}P_{l}(\hat{p}_1\.\hat{p}_2)=-i\gamma \lim_{\e_{+}\rightarrow 0}\frac{\Gamma(1+i\gamma)}{\Gamma(1-i\gamma)}\Bigg(\frac{2}{(1-\hat{p}_1\.\hat{p}_2)}\Bigg)^{1+i\gamma-\e_{+}},\label{thesum}
\end{gather}
where the equality (\ref{thesum}) should be understood in the distributional sense, meaning that both sides yield the same result when integrated against a test function \cite{Taylor,herbst1974connectedness}. A straightforward way to verify this is to project both sides onto $P_{l'}$ by integrating over the relative angle, for example using $\int \mathrm{d}^2\Omega_{\hat{p}_1}\,P_{l'}(\hat{p}_1 \cdot \hat{p}_2)(\cdots)$, and check that they produce the same result (see (\ref{used Gradshteyn}) for the relevant integral evaluation). The positive infinitesimal exponent in $\e_{+}>0$ was shown by Herbst \cite{herbst1974connectedness} to be the unique distributional interpretation of the Coulomb amplitude consistent with unitarity when including the forward direction. A more detailed discussion of these points is provided in the next subsection.

Using (\ref{thesum}) in (\ref{almost done}), and converting the Dirac delta function's argument to $E=\frac{|\vec{p}|^2}{2m}$, we conclude that
\begin{align}
    A^{\text{NRQM}}(\vec{p}_1,\vec{p}_2) &= \int \mathrm{d}^3 x\, \phi^{\star}_{\text{out}}(\vec{x},\vec{p}_2)\phi_{\text{in}}(\vec{x},\vec{p}_1) \\
    &=-\frac{i\pi}{2}\frac{e_1e_2}{|\vec{p}|^2}\delta(E_1-E_2)\frac{\Gamma(1+i\gamma)}{\Gamma(1-i\gamma)}\Bigg(\frac{2}{(1-\hat{p}_1\.\hat{p}_2)}\Bigg)^{1+i\gamma},\quad \hat{p}_1\.\hat{p}_2\neq 1
\end{align}
which can be seen to be equal to (\ref{ampNRQMint}) upon using $(1-\hat{p}_1\.\hat{p}_2)=1-\cos\theta=2\sin^2(\theta/2)$.

\subsubsection*{The distributional nature of the amplitude}
The distributional equality (\ref{thesum}) is a subtle point, so we summarize the key arguments from \cite{Taylor,herbst1974connectedness}. First, the large-$l$ behavior of the left-hand side of (\ref{thesum}),
\begin{gather} 
\lim_{l\rightarrow \infty} e^{2i\delta_l} = l^{2i\gamma} \big( 1 + \mathcal{O}(1/l) \big), 
\end{gather}
indicates that the partial-wave sum diverges. However, as demonstrated in \cite{Taylor}, this sum defines a perfectly well-behaved distribution that acts unambiguously and unitarily on any sufficiently smooth wavepacket. We now explain this in more detail.

Scattering amplitudes are distributions, so the fundamental requirement for an amplitude \( A(p_{\text{out}}, p_{\text{in}}) \) is that it provides a well-defined and unitary mapping from in-wavepackets \( f_{\text{in}}(\vec{p}_{\text{in}}) \) to out-wavepackets \( f_{\text{out}}(\vec{p}_{\text{out}}) \):  
\begin{gather}
    f_{\t{out}}(\vec{p}_{\t{out}})=\int\frac{\DD^3p_{\t{in}}}{(2\pi)^3} A(p_{\t{out}},p_{\t{in}})f_{\t{in}}(\vec{p}_{\t{in}})
\end{gather}
Let us consider an in-wavepacket that can be decomposed in the partial-wave basis,  
\begin{gather}
    f_{\text{in}}(\vec{p}_{\text{in}}) = \sum_{l=0}^{\infty} f^{\text{in}}_l(|\vec{p}_{\text{in}}|) P_l(\hat{p}_{\text{in}} \cdot \hat{z}),\label{in wavepacket}
\end{gather}
where \( f^{\text{in}}_l \) are the partial-wave coefficients. For simplicity, we have assumed a wavepacket that is symmetric about the \( \hat{z} \)-axis, though this assumption is not essential.

If we take the amplitude to be (\ref{almost done})
\begin{gather}
   A(\hat{p}_{\t{2}},\hat{p}_{\t{1}})= \delta(|\vec{p}_1|-|\vec{p}_2|) \frac{2\pi^2}{|\vec{p}_1|^2}\sum_{l}(2l+1)\frac{\Gamma(1+l+i\gamma)}{\Gamma(1+l-i\gamma)}P_{l}(\hat{p}_1\.\hat{p}_2)\label{eq:action in partial}
\end{gather}
then we can examine the action of the amplitude on the in-wavepacket
\begin{align}
 \int\frac{\DD^3p_{\t{in}}}{(2\pi)^3} A(p_{\t{out}},p_{\t{in}})f_{\t{in}}(\vec{p}_{\t{in}})&= \sum_{l=0}^{\infty}\int\frac{\DD^3p_{\t{in}}}{(2\pi)^3} A(p_{\t{out}},p_{\t{in}})f^{\t{in}}_l(|\vec{p}_{\t{in}}|) P_l(\hat{p}_{\t{in}}\.\hat{z})\\
 &=\sum_{l=0}^{\infty}f^{\t{in}}_l(|\vec{p}_{\t{out}}|)e^{2i\delta_l}P_{l}(\hat{p}_{\t{out}}\.\hat{z})\label{eq:which functions}
\end{align}
where in going to (\ref{eq:which functions}) we used (\ref{angular sum of legendres}). Thus, (\ref{eq:action in partial}) defines a well-defined distribution acting on wavepackets, provided that the sum (\ref{eq:which functions}) converges. Furthermore, the action is unitary since each partial-wave coefficient \( f_{l}^{\text{in}} \) is only modified by a phase, ensuring that the norm of the wavepacket is preserved under the action of the amplitude.

Next, we need to show that the amplitude  
\begin{gather}
   \tilde{A}(\hat{p}_{2},\hat{p}_{1})= -i\gamma \lim_{\e_{+}\rightarrow 0}\delta(|\vec{p}_1|-|\vec{p}_2|) \frac{2\pi^2}{|\vec{p}_1|^2}\frac{\Gamma(1+i\gamma)}{\Gamma(1-i\gamma)}\Bigg(\frac{2}{(1-\hat{p}_1\.\hat{p}_2)}\Bigg)^{1+i\gamma-\e_{+}}\label{atilde}
\end{gather}
has exactly the same distributional action as (\ref{eq:which functions}). Acting on the in-wavepacket with (\ref{atilde}),
\begin{align*}
    &\int\frac{\DD^3p_{\t{in}}}{(2\pi)^3} \tilde{A}(p_{\t{out}},p_{\t{in}})f_{\t{in}}(\vec{p}_{\t{in}})\\
    &=-\frac{i\gamma}{4\pi} \frac{\Gamma(1+i\gamma)}{\Gamma(1-i\gamma)}\int\DD^2\Omega_{\t{in}}\Bigg(\frac{2}{(1-\hat{p}_{\t{in}}\.\hat{p}_{\t{out}})}\Bigg)^{1+i\gamma-\e_{+}}f_{\t{in}}(\vec{p}_{\t{in}})\numberthis\\
    &=-\frac{i\gamma}{4\pi} \frac{\Gamma(1+i\gamma)}{\Gamma(1-i\gamma)}\int\DD^2\Omega_{\t{in}}\int_{-1}^1\DD x \delta(x-\hat{p}_{\t{in}}\.\hat{p}_{\t{out}})\Bigg(\frac{2}{1-x}\Bigg)^{1+i\gamma-\e_{+}}f_{\t{in}}(\vec{p}_{\t{in}})\numberthis\\
    &=-\frac{i\gamma}{4\pi} \frac{\Gamma(1+i\gamma)}{\Gamma(1-i\gamma)}\int\DD^2\Omega_{\t{in}}\int_{-1}^1\DD x\sum_{l'} \frac{2l'+1}{2}P_{l'}(x)P_{l'}(\hat{p}_{\t{in}}\.\hat{p}_{\t{out}})\Bigg(\frac{2}{1-x}\Bigg)^{1+i\gamma-\e_{+}}f_{\t{in}}(\vec{p}_{\t{in}})\numberthis\label{introduce delta}\\
    &=-\frac{i\gamma}{2} \frac{\Gamma(1+i\gamma)}{\Gamma(1-i\gamma)}\sum_l f^{\t{in}}_l(|\vec{p}_{\t{out}}|)P_{l}(\hat{p}_{\t{out}}\.\hat{z})\int_{-1}^1\DD x\Bigg(\frac{2}{1-x}\Bigg)^{1+i\gamma-\e_{+}}P_l(x).\numberthis\label{intermidairy and still going}
\end{align*}
In going to (\ref{introduce delta}), we used the identity  
\begin{gather}
    \delta(x - y) = \sum_{l=0}^{\infty} \frac{2l+1}{2} P_l(x) P_l(y),
\end{gather}  
which expresses the Dirac delta function as a sum over Legendre polynomials.  In going to the equality (\ref{intermidairy and still going}), we used the partial-wave expansion of the in-wavepacket \( f_{\text{in}}(\vec{p}_{\text{in}}) \) given in (\ref{in wavepacket}), and then applied (\ref{angular sum of legendres}) to perform the angular integral. Next we perform the remaining $x$-integral,
\begin{align*}
    &\int\frac{\DD^3p_{\t{in}}}{(2\pi)^3} \tilde{A}(p_{\t{out}},p_{\t{in}})f_{\t{in}}(\vec{p}_{\t{in}})\\
    &=-\frac{i\gamma}{2} \frac{\Gamma(1+i\gamma)}{\Gamma(1-i\gamma)}\sum_l f^{\t{in}}_l(|\vec{p}_{\t{out}}|)P_{l}(\hat{p}_{\t{out}}\.\hat{z})\int_{-1}^1\DD x\Bigg(\frac{2}{1-x}\Bigg)^{1+i\gamma-\e_{+}}P_l(x)\numberthis\\
    &=-i\gamma\frac{\Gamma(1+i\gamma)}{\Gamma(1-i\gamma)}\sum_l f^{\t{in}}_l(|\vec{p}_{\t{out}}|)P_{l}(\hat{p}_{\t{out}}\.\hat{z})\frac{(-1)^l[\Gamma(-i\gamma+\e_{+})]^2}{\Gamma(1-i\gamma+l+\e_{+})\Gamma(-l-i\gamma+\e_{+})}\numberthis\label{used Gradshteyn}\\
    &=\sum_{l=0}^{\infty}f^{\t{in}}_l(|\vec{p}_{\t{out}}|)e^{2i\delta_l}P_{l}(\hat{p}_{\t{out}}\.\hat{z}).\numberthis\label{final action}
\end{align*}
In going to (\ref{used Gradshteyn}), we applied Eq. (7.127) from \cite{gradshteyn2014table}. Next, in transitioning to (\ref{final action}), we analytically continued to \( \e_{+} \to 0 \) and then used Euler's reflection formula to simplify the expression. This confirms that (\ref{final action}) and (\ref{eq:which functions}) produce identical results, thereby validating the distributional equality (\ref{thesum}). Moreover, this establishes that the amplitude has a well-defined, unambiguous, and unitary (norm-preserving) action on the space of wavefunctions for which the sum (\ref{final action}) converges.

The $\e_{+}$ exponent is an especially natural regulator in light of the fact that when one allows for soft-photon production, the exponent of the amplitude would naturally acquire a finite damping factor, namely the $R$ in (\ref{eq:Weinbergexponential}).
\subsection{NLO amplitude integral}\label{app:NLO integral}
In this appendix, we compute the integral for the NLO amplitude (\ref{exp})
\begin{equation}
    I(p_1,p_2)=\delta(E_1-E_2)\int\DD^3 y \,\, \phi_{\text{out}}^{\star}(y,p_2) \frac{1}{|\vec{y}|^2}\phi_{\text{in}}(y,p_1) .
\end{equation}
Expanding in terms of spherical harmonics and working in the rest frame of the source particle as in appendix \ref{app:LO Int}, we obtain
\begin{align}
    I(p_1,p_2)=&\delta(E_1-E_2)e^{-\pi \gamma}4\pi\sum_{l=0}^{\infty}\frac{1}{2l+1}P_{l}(\hat{p}_1\.\hat{p}_2)\Big[\frac{\Gamma(1+l+i\gamma)}{(2l)!}\Big]^2(-1)^l(2ip)^{2l}f(p,\gamma,l)\label{eq:I}\\
   f(p,\gamma,l)\defined &\int_0^{\infty}\DD r\,\, r^{2l}{}_1F_1\Big(1+l-i\gamma,2l+2,2ipr\Big){}_1F_1\Big(1+l+i\gamma,2l+2,-2ipr\Big)
\end{align}
Here $p=|\vec{p}_1|=|\vec{p}_2|$ and $\gamma=\gamma_1=\gamma_2$ on the support of the delta function.
To evaluate $f(p,\gamma,l)$ we will use the integral representation of the confluent hypergeometric function
\begin{gather}
    {}_1F_1(a, b, z) = \frac{\Gamma(b)}{\Gamma(a)\Gamma(b-a)} \int_{0}^{1} e^{zt} t^{a-1} (1-t)^{b-a-1} \, \DD t,\qquad \text{Re}[b]>\text{Re}[a]>0.
\end{gather}
So we have
\begin{align}
    &f(p,\gamma,l)=\frac{[\Gamma(2l+2)]^2}{|\Gamma(1+l+i\gamma)|^4}  \int_0^1 \DD s\int_0^1\DD t\int_0^{\infty}\DD r\,\, r^{2l}e^{2ipr (t-s+i\e_{+})}t^{l-i\gamma}(1-t)^{l+i\gamma}s^{l+i\gamma}(1-s)^{l-i\gamma}\label{eq:regulate}\\
    &=\frac{1}{(2ip)^{2l+1}}\frac{[\Gamma(2l+2)]^2\Gamma(2l+1)}{|\Gamma(1+l+i\gamma)|^4}  \int_0^1 \DD s \int_0^1\DD t \,\, \frac{t^{l-i\gamma}(1-t)^{l+i\gamma}s^{-l-1+i\gamma}(1-s)^{l-i\gamma}}{(1-\frac{t}{s-i\e_{+}})^{2l+1}}\\
    &=\frac{1}{(2ip)^{2l+1}}\frac{\Gamma(2l+2)\Gamma(2l+1)}{|\Gamma(1+l+i\gamma)|^2}  \int_0^1 \DD s \,\, s^{-l-1+i\gamma}(1-s)^{l-i\gamma}{}_2F_1\Big(1+2l,1+l-i\gamma,2l+2,\frac{1}{s-i\e_{+}}\Big)
\end{align}
where we added a positive infinitesimal $\e_{+}$ to give a physical analytic continuation of the integral (\ref{eq:regulate}), which we will take to zero at the end of the calculation. We now use the following hypergeometric function identity to perform the analytic continuation
\begin{align*}
{}_2F_1(a, b, c, z) = &\frac{\Gamma(b - a) \Gamma(c)}{\Gamma(c - a) \Gamma(b)} \frac{{}_2F_1(a, a + 1 - c, a + 1 - b, 1/z)}{(-z)^a} \\
+&\frac{\Gamma(a - b) \Gamma(c)}{\Gamma(c - b) \Gamma(a)} \frac{{}_2F_1(b + 1 - c, b, b + 1 - a, 1/z)}{(-z)^b},\qquad z \notin (0,1)\numberthis
\end{align*}
which for our case gives us various factors of $(-\frac{1}{s- i\e_{+}})^p=\frac{1}{s^p}(e^{-i\pi p})$. In our case we obtain
\begin{align*}
    &{}_2F_1\Big(1+2l,1+l-i\gamma,2l+2,\frac{1}{s-i\e_{+}}\Big)=\\
    &- \frac{s^{2 l+1} \Gamma\left(2l+2\right) \Gamma\left(-l - i \gamma\right)}{\Gamma\left(1 + l - i \gamma\right)}+
(-1)^{l} e^{\pi \gamma} (2l+1) s^{2l+1} \text{Beta}\left[s - i \e_{+}, -l - i \gamma, -l + i \gamma\right]\numberthis
\end{align*}
This then gives us two terms for $f(p,\gamma,l)$
\begin{gather}
    f(p,\gamma,l)=\text{H}_1+\text{H}_2
    \end{gather}
    where the first term is
    \begin{align}
    \text{H}_1&=-\frac{1}{(2ip)^{2l+1}}\frac{\Gamma(2l+2)\Gamma(2l+1)}{|\Gamma(1+l+i\gamma)|^2} \frac{ \Gamma\left(2l+2\right) \Gamma\left(-l - i \gamma\right)}{\Gamma\left(1 + l - i \gamma\right)} \int_0^1 \DD s\,\, s^{l+i\gamma}(1-s)^{l-i\gamma}\\
    &=- \frac{1}{(2 i p)^{1 + 2 l}}\frac{ \Gamma(1 + 2 l) \Gamma(2 + 2 l) \Gamma(-l - i \gamma)}{\Gamma(1 + l - i \gamma)}
\end{align}
and the second term is
\begin{align}
    \text{H}_2&=\frac{(-1)^{l}}{(2ip)^{2l+1}}\frac{\Gamma(2l+2)\Gamma(2l+1)}{|\Gamma(1+l+i\gamma)|^2}  e^{\pi \gamma} (2l+1) \int_0^1\DD s s^{l+i\gamma}(1-s)^{l-i\gamma} \text{Beta}\left[s - i \e_{+}, -l - i \gamma, -l + i \gamma\right]\\
    &=\frac{(-1)^{l}}{(2ip)^{2l+1}}\frac{\Gamma(2l+2)\Gamma(2l+1)}{|\Gamma(1+l+i\gamma)|^2}  e^{\pi \gamma}  \Big(H_{l-i\gamma}-H_{-1-l-i\gamma} \Big)
\end{align}
where $H_n$ is the $n^{th}$ harmonic number. Thus far we have established
\begin{align*}
      &f(p,\gamma,l)\defined \int_0^{\infty}\DD r\,\, r^{2l}{}_1F_1\Big(1+l-i\gamma,2l+2,2ipr\Big){}_1F_1\Big(1+l+i\gamma,2l+2,-2ipr\Big)\\
& =\frac{1}{(2ip)^{2l+1}}\frac{\Gamma(2l+1)\Gamma(2l+2)}{|\Gamma(1+l-i\gamma)|^2}\Bigg(-\Gamma(1+l+i\gamma)\Gamma(-l-i\gamma)+(-1)^le^{\pi\gamma}\Big(H_{l-i\gamma}-H_{-1-l-i\gamma} \Big)\Bigg) \numberthis    \label{eq:intermediate}
\end{align*}
We would now like to bring this to a form that can be compared to the exact result; see the order-$\alpha^2$ term in (\ref{eq:exactexpansion}). Using the following formulas, which involve the Euler-Mascheroni constant $\gamma_{E}$ and the digamma function $\psi(x)$,
\begin{gather}
    H_{n}=\psi(n+1)+\gamma_{E}\\
    \psi(1-x)-\psi(x)=\pi\cot\pi x
\end{gather}
we can write
\begin{align}
    H_{-1-l-i\gamma}=H_{l+i\gamma}+\pi\cot\pi(l+i\gamma)\label{expressharmonic}
\end{align}
This $\pi\cot\pi(l+i\gamma)$ term then combines with the first term in (\ref{eq:intermediate}) in the following way 
\begin{align*}
    &(-1)^l(-1)e^{\pi\gamma}\pi\cot\pi(l+i\gamma)-\Gamma(-l - i \gamma)\Gamma(1+l+i\gamma)\\
    &=(-1)^l(-1)e^{\pi\gamma}\pi\cot\pi(i\gamma)-\frac{\pi}{\sin\pi(1+l+i\gamma)}\numberthis\\
    &=(-1)^li\pi e^{\pi\gamma}\numberthis\label{combineterms}
\end{align*}
where we used Euler's reflection formula in the first equality, and then expressed all trigonometric functions in terms of exponentials to deduce the final equality. Returning to (\ref{eq:I}) with our value for $f(p,\gamma,l)$ given by (\ref{eq:intermediate}) while also using the results (\ref{expressharmonic}, \ref{combineterms}), we have that
\begin{gather}
     I(p_1,p_2)=\delta(E_1-E_2)\frac{2\pi}{(i|\vec{p}|)} \sum_{l=0}^{\infty}\frac{\Gamma(1+l+i\gamma)}{\Gamma(1+l-i\gamma)}  P_{l}(\hat{p}_1\.\hat{p}_2) \Big(H_{l-i\gamma}-H_{l+i\gamma} +i\pi\Big).
\end{gather}

\subsection{NLO summation}\label{app:NLO summation}
In this section we perform the summation
\begin{align}
    f(\theta,\gamma)&=\sum_{l=0}^{\infty}\frac{\Gamma(1+l+i\gamma)}{\Gamma(1+l-i\gamma)}  P_{l}(\hat{p}_1\.\hat{p}_2) \Big(H_{l-i\gamma}-H_{l+i\gamma} +i\pi\Big)\\
    &\defined f_1(\theta,\gamma)+f_{2}(\theta,\gamma)
\end{align}
where
\begin{align}
    f_1(\theta,\gamma)&=i\pi\sum_{l=0}^{\infty}\frac{\Gamma(1+l+i\gamma)}{\Gamma(1+l-i\gamma)}  P_{l}(\hat{p}_1\.\hat{p}_2) \\
      f_2(\theta,\gamma)&=\sum_{l=0}^{\infty}\frac{\Gamma(1+l+i\gamma)}{\Gamma(1+l-i\gamma)}  P_{l}(\hat{p}_1\.\hat{p}_2) \Big(H_{l-i\gamma}-H_{l+i\gamma}\Big).
\end{align}
The series is non-convergent. We interpret it distributionally: the series and the function must agree when projected onto $P_l(\cos\theta)$ \cite{Taylor}; i.e., we seek a function that obeys
\begin{gather}
    \int_{-1}^1\DD (\cos\theta) P_{l}(\cos\theta)f(\theta,\gamma)=\frac{2}{2l+1}\frac{\Gamma(1+l+i\gamma)}{\Gamma(1+l-i\gamma)}  \Big(H_{l-i\gamma}-H_{l+i\gamma} +i\pi\Big).
\end{gather}
We will evaluate $f_1$ in closed form, but are only able to bring $f_2$ into the form of an integral that is useful for obtaining IR-finite results.
We can compute $f_1$ as follows
\begin{align}
    f_1(\theta,\gamma)&=\lim_{\text{Im}[\gamma]\rightarrow 0^{+}}\frac{i\pi}{\Gamma(-2i\gamma)}\sum_l\int_0^1\DD x\,\, x^{l+i\gamma}(1-x)^{-2i\gamma-1}P_{l}(\cos\theta)\\
    &=\lim_{\text{Im}[\gamma]\rightarrow 0^{+}}\frac{i\pi}{\Gamma(-2i\gamma)}\int_0^{1}\DD x \frac{x^{i\gamma}(1-x)^{-2i\gamma-1}}{\sqrt{x^2-2x\cos\theta+1}}
\end{align}
where in the first line we used an integral representation for the Beta function, giving the anomalous dimension an infinitesimal positive imaginary part to make the integral well-defined. In the second line we used the generating-function formula for the Legendre polynomials in order to perform the sum over $l$. We then make the change of variables
\begin{gather}
    x=\Big(\sqrt{1+w^2}-w\Big)^2
\end{gather}
and find
\begin{align*}
     f_1(\theta,\gamma)&=i\pi\lim_{\text{Im}[\gamma]\rightarrow 0^{+}}\frac{2^{-1 - 2i \gamma}}{\Gamma(-2i\gamma)}\int_0^{\infty}\DD w\frac{ w^{-1 - 2i \gamma} }{\sqrt{1 + w^2} \sqrt{w^2 + \sin^2\left(\frac{\theta}{2}\right)}}\numberthis\\
    &=\frac{i\pi}{2}\Bigg(\frac{\Gamma\left(\frac{1}{2} + i \gamma\right) \,}{ \Gamma\left(\frac{1}{2} - i \gamma\right)}\csc\left(\frac{\theta}{2}\right)^{1 + 2i \gamma} {}_2F_1\left(\frac{1}{2}, -i \gamma, \frac{1}{2} - i \gamma, \sin^2\frac{\theta}{2}\right)\label{NLOclosedformfirst}\\ &\qquad +\frac{\Gamma\left(-\frac{1}{2} - i \gamma\right) \Gamma\left(1 + i \gamma\right) }{2^{1 + 2i \gamma}\sqrt{\pi} \Gamma\left(-2i \gamma\right)}\,\,{}_2F_1\left(\frac{1}{2}, 1 + i \gamma, \frac{3}{2} + i \gamma, \sin^2\frac{\theta}{2}\right)\Bigg) .\numberthis
\end{align*}
We can perform the sum in $f_2$ by using
\begin{gather}
        \frac{\Gamma(1+l+i\gamma)}{\Gamma(1+l-i\gamma)}\Big(H_{l-i\gamma}-H_{l+i\gamma}\Big)=- \frac{1}{\Gamma(-2i\gamma)} \int_{0}^{1} x^{l + i \gamma} \left(1 - x\right)^{-2i \gamma - 1} \log(x) \, dx,\qquad \text{Im}[\gamma]>0
\end{gather}
Performing the same steps as before we arrive at
\begin{align}
     f_2(\theta,\gamma)&=\frac{2^{ - 2i \gamma}}{\Gamma(-2i\gamma)}\int_0^{\infty}\DD w\frac{ w^{-1 - 2i \gamma} \text{arcsinh}(w)}{\sqrt{1 + w^2} \sqrt{w^2 + \sin^2\left(\frac{\theta}{2}\right)}}\\
     &=\frac{2^{ - 2i \gamma}}{\Gamma(-2i\gamma)}\int_0^{\infty}\DD \psi \frac{\psi\sinh(\psi)^{-1 - 2i \gamma}}{\sqrt{\sinh^2(\psi) + \sin^2\left(\frac{\theta}{2}\right)}}\label{NLO2 int}
\end{align} 
where in the last line we made the change of variable $w=\sinh\psi$. We have only been able to evaluate this integral in terms of derivatives of ${}_3F_2$ functions, which are no more enlightening than the integral representation (\ref{NLO2 int}), which itself can be used to generate numerical values for the amplitude.
\section{The relativistic Coulomb Green's function}\label{app:Greensfunction}
In appendix \ref{app:boundstates} we give the explicit expression for the on-shell $\phi_{nlm}$ and off-shell $\chi_{nlm}$ bound states, as well as the off-shell representation of the ``semi-free'' Green's function. In appendix \ref{app:proveGreens} we verify that the off-shell representation of the Green's function
\begin{align*}
    G^{\pm}(x,y)=&-\int\frac{\DD^4p}{(2\pi)^4} \,\, \frac{\chi_{\text{in/out}}(x,p,u_s)\chi_{\text{in/out}}^{\star}(y,p,u_s)}{p^2-m^2\pm i\e}\\
    &+\frac{m^2}{\pi}\sum_{l=0}^{\infty}\sum_{n=l+1}^{\infty}\sum_{m=-l}^l\int_{-\infty}^{\infty}\DD \omega\theta(-\alpha\omega)\frac{\chi_{nlm}(\omega,x)\chi^{\star}_{nlm}(\omega,y)}{m^2-\Big(1+\frac{\alpha^2}{n^2}\Big)\omega^2},\numberthis\label{offshellgreens}
\end{align*}
satisfies the Green's function equation
\begin{equation}
    (\Box +m^2+2ie_1A^{\mu}\d_{\mu})_xG^{\pm}(x,y)=\delta^{4}(x-y)
\end{equation}
where $\chi$ are the off-shell wavefunctions, i.e. (\ref{eq:instate}, \ref{eq:outstate}) where we do not impose $p^2=m^2$. In appendix \ref{app:boundstatecancellation} we analyse the poles of the integrand (\ref{offshellgreens}) in the complex $p^0$-plane. We then perform the integral over $p^0$, demonstrating a cancellation between some of these poles that ensures that retarded/advanced boundary conditions are properly implemented. We then deduce the purely on-shell form of the Green's function (\ref{eq:onshellgreens}) used in the paper. In the final subsection, appendix \ref{app:vanishretardedgreens}, we demonstrate that the retarded Green's function vanishes at equal times; we use this result when computing the NLO amplitude at (\ref{usethisidentity}).
\subsection{Bound-state wavefunctions}\label{app:boundstates}
The bound-state wavefunctions for the semi-free KG equation
\begin{gather}
        \Bigg(\Box +m^2+2i\frac{\alpha}{r} \d_{t} \Bigg)\phi_{nlm}(x)=0,
\end{gather}
are
\begin{gather}
    \phi_{nlm}(x)=\theta(-\alpha E_n)N(n,l)e^{-iE_nt}Y_{lm}(\theta,\phi)R_{nl}(r)
\end{gather}
where
\begin{gather}
     R_{nl}(r)=(2\eta_n r)^le^{-\eta_n r}{}_1F_1(1+l+n,2l+2,2\eta_n r)\\
    \eta_n=\frac{\alpha}{n}E_{n},\qquad E_n=-\text{sign}(\alpha) \frac{m}{\sqrt{1+\frac{\alpha^2}{n^2}}},\qquad N(n,l)=\frac{(-\alpha E_n)^{\frac{3}{2}}}{m n^2}\sqrt{2\frac{(l+n)!}{(n-l-1)!}}\frac{1}{(2l+1)!}
    \\ n=l+1,l+2,\ldots\qquad\quad  l=0,1,2,\ldots
\end{gather}
where the normalization is chosen so that 
\begin{gather}
    \braket{\phi_{nlm}|\phi_{n'l'm'}}_{\text{SQED}}=\frac{1}{E_n}\delta_{nn'}\delta_{ll'}\delta_{mm'}.\label{boundortho}
\end{gather}
To construct the Green's function we require the off-shell solutions
\begin{gather}
    \Bigg(\Box +m^2+2i\frac{\alpha}{r} \d_{t} \Bigg)\chi_{nlm}(\omega,x)=\Bigg(m^2-\omega^2\Big(1+\frac{\alpha^2}{n^2}\Big)\Bigg)\chi_{nlm}(\omega,x)\label{eigenvalue}
\end{gather}
which is solved by taking the energy off-shell,
\begin{gather}
     \chi_{nlm}(\omega,x)=\phi_{nlm}(x)\at{E_n=\omega},
\end{gather}
where $\omega$ is unrestricted.
Thus, as we will prove shortly, the bound-state contribution to the Green's function is
\begin{gather}
        G_{B}(x,y)=\frac{m^2}{\pi}\sum_{l=0}^{\infty}\sum_{n=l+1}^{\infty}\sum_{m=-l}^l\int_{-\infty}^{\infty}\DD \omega\theta(-\alpha\omega)\frac{\chi_{nlm}(\omega,x)\chi^{\star}_{nlm}(\omega,y)}{m^2-\Big(1+\frac{\alpha^2}{n^2}\Big)\omega^2}.\label{boundstate contibution}
\end{gather}
\subsection{Green's function equation}\label{app:proveGreens}
In this section we prove that the sum of the continuum and bound-state Green's functions
\begin{align}
    G(x,y)&=G_C(x,y)+G_B(x,y)\\
    G_C(x,y)&=-\int\frac{\DD^4p}{(2\pi)^4} \,\, \frac{\chi_{\text{in/out}}(x,p,u_s)\chi_{\text{in/out}}^{\star}(y,p,u_s)}{p^2-m^2\pm i\e}\label{continiuumgreens}
\end{align}
satisfies the Green's function equation for the semi-free KG equation
\begin{gather}
    \Big(\Box+m^2+2ie_1A^{\mu}\d_{\mu}\Big)G(x,y)=\delta^4(x-y)\,.\label{toprove}
\end{gather}
Mukunda \cite{mukunda1978completeness} has proven the following completeness relation for the non-relativistic Coulomb radial wavefunctions
\begin{equation}
\theta(-a)\sum_{n=l+1}^{\infty} \rho_{nl}(r,a) \rho^{\star}_{nl}(r',a) + \int_0^{\infty} v_{l}(k,r,a) v^{\star}_{l}(k,r',a) dk = r^{-2} \delta(r - r').\label{complete Makunda}
\end{equation}
where $v_l(k,r,a)$ and $\rho_{nl}(r,a)$ are the non-relativistic continuum and bound-state radial wavefunctions, respectively\footnote{When comparing to \cite{mukunda1978completeness} one has to replace $a\rightarrow -a$ due to different sign conventions for the coupling, and subsequently use the relation ${}_1F_1(a,b,-z)=e^{-z}{}_1F_1(b-a,b,z)$.}
\begin{gather}
v_{l}(k,r,a) = \sqrt{\frac{2}{\pi}}\frac{k}{(2l + 1)!}  e^{-\frac{\pi }{2ak}} \Bigg|\Gamma\left(l + 1 + \frac{i}{ak}\right)\Bigg|  (2kr)^l e^{-ikr} {}_1F_1\left(l + 1 - \frac{i}{ak}, 2l + 2; 2ikr\right). \label{eq:contradial}\\
\rho_{nl}(r,a) =  \frac{2 }{n^2 (-a)^{3/2}(2l + 1)!}  
 \sqrt{\frac{(n + l)!}{(n - l - 1)!}}  \left( \frac{2r}{na} \right)^l e^{-\frac{r}{na}} {}_1F_1\Bigg(l + 1 + n, 2l + 2, \frac{2r}{na}\Bigg) \label{boundradial}
\end{gather}
where in NRQM $a=\frac{1}{me^2}$, and the $\theta(-a)$ in (\ref{complete Makunda}) indicates that the bound states only contribute to the completeness relation for attractive potentials. In order to anticipate some cancellations in this section, it is useful to note that the last three radially dependent terms in the continuum and bound-state wavefunctions (\ref{eq:contradial}, \ref{boundradial}) are the same under the replacement $k\rightarrow -\frac{i}{na}$. Fortunately for us, the \textit{on-shell} NRQM bound-state wavefunctions $\rho_{nl}$ are identical to the \textit{off-shell} relativistic wavefunctions upon making the replacement $a\rightarrow \frac{1}{\alpha \omega}$
\begin{gather}
    \rho_{nl}\Big(r,\frac{1}{\alpha\omega}\Big)=m\sqrt{2}N(n,l)R_{nl}(r)\at{E_n=\omega}\label{boundstatesintermsofNRQM}
\end{gather}
We can also express the off-shell relativistic continuum wavefunctions in terms of the on-shell NRQM continuum wavefunctions
\begin{align}
    \chi_{\text{in}}(x,p,u_s^r)&=e^{-\frac{\pi\gamma}{2}}e^{-i\omega t}\sum_{l=0}^{\infty}\frac{\Gamma(1+l+i\gamma)}{(2l)!}(2i|\vec{p}|r)^le^{-i|\vec{p}|r}{}_1F_1\Big(1+l-i\gamma,2l+2,2i|\vec{p}|r\Big)P_{l}(\hat{x}\.\hat{p})\\
     &=e^{-i\omega t}\frac{1}{|\vec{p}|}\sqrt{\frac{\pi}{2}}\sum_{l=0}^{\infty}v_{l}\Big(p,r,\frac{1}{\alpha\omega}\Big)i^l(2l+1)e^{i\delta_l}P_{l}(\hat{x}\.\hat{p})\label{offshellcontNRQM}
\end{align}
where $e^{i\delta_l}$ is the phase
\begin{gather}
    e^{i\delta_l}=\sqrt{\frac{\Gamma(1+l+i\gamma)}{\Gamma(1+l-i\gamma)}}
\end{gather}
In the following steps we will perform the angular integration relevant to the continuum contribution to the Green's function using
\begin{align}
    \int\DD^2\Omega_p P_l(\hat{x}_1\.\hat{p})P_{l'}(\hat{x}_2\.\hat{p})&=\delta_{ll'}\frac{4\pi}{2l+1}P_{l}(\hat{x}_1\.\hat{x}_2)  \\
    &=\delta_{ll'}\frac{16\pi^2}{(2l+1)^2} \sum_{m=-l}^lY_l^{m}(\hat{x}_1)\Big( Y_l^{m}(\hat{x}_2)\Big)^{\star}.\label{angularidentity}
\end{align}
We now calculate the action of the semi-free KG differential operator on the proposed Green's function
\begin{align*}
&\Big(\Box+m^2+2ie_1A^{\mu}\d_{\mu}\Big)\Big(G_B(x,y)+G_C(x,y)\Big)\numberthis\\
    &=\frac{m^2}{\pi}\sum_{l=0}^{\infty}\sum_{n=l+1}^{\infty}\sum_{m=-l}^l\int_{-\infty}^{\infty}\DD p^0\theta(-\alpha p^0)\chi_{nlm}(x,p^0)\chi^{\star}_{nlm}(y,p^0) +\int\frac{\DD^4p}{(2\pi)^4} \,\, \chi_{\text{in/out}}(x,p)\chi_{\text{in/out}}^{\star}(y,p)\numberthis \label{nodenom}\\
    &=\frac{1}{2\pi}\sum_{l=0}^{\infty}\sum_{m=-l}^lY_l^{m}(\hat{x})\Big( Y_l^{m}(\hat{y})\Big)^{\star}\int_{-\infty}^{\infty}\DD p^0 e^{-ip^0(x^0-y^0)}\\
    &\times\Bigg[\sum_{n=l+1}^{\infty} \theta(-\alpha p^0)\rho_{nl}\Big(r_x,\frac{1}{\alpha p^0}\Big)\rho^{\star}_{nl}\Big(r_y,\frac{1}{\alpha p^0}\Big)
    +\int_{0}^{\infty}\DD |\vec{p}| \,\, v_l\Big(|\vec{p}|,r_x,\frac{1}{\alpha p^0}\Big)v^{\star}_l\Big(|\vec{p}|,r_y,\frac{1}{\alpha p^0}\Big)\Bigg]\numberthis\label{lotsasteps}\\
    &=\frac{1}{2\pi}\sum_{l=0}^{\infty}\sum_{m=-l}^lY_l^{m}(\hat{x})\Big( Y_l^{m}(\hat{y})\Big)^{\star}\int_{-\infty}^{\infty}\DD p^0 e^{-ip^0(x^0-y^0)}\frac{\delta(r_x-r_y)}{r_x^2}\numberthis\label{complete}\\
    &=\delta^4(x-y).\numberthis
\end{align*}
In going to (\ref{nodenom}) we used the eigenvalue equations (\ref{eigenvalue}, \ref{eigenvaluecontiniuum}) of the off-shell wavefunctions, which cancelled the eigenvalue denominator factors. In going to (\ref{lotsasteps}) we expressed the off-shell bound-state wavefunctions in terms of the NRQM radial wavefunctions (\ref{boundstatesintermsofNRQM}) and spherical harmonics. We also expressed the off-shell continuum wavefunctions in terms of the NRQM radial wavefunctions (\ref{offshellcontNRQM}) and spherical harmonics, and then performed the angular integral over momentum using (\ref{angularidentity})\footnote{The in- and out-state expansions in terms of the continuum NRQM radial wavefunctions (\ref{offshellcontNRQM}) differ by phases $e^{\pm i\delta_l}$, which cancel against their complex conjugates in (\ref{lotsasteps}). This shows that we can use either the in- or out-state wavefunctions to express the Green's function.} To obtain (\ref{complete}) we used Mukunda's completeness relation (\ref{complete Makunda}), and in going to the last line we used the orthogonality of the spherical harmonics
\begin{equation}
\sum_{l,m} Y_{lm}(\theta, \phi) Y^*_{lm}(\theta', \phi') = \delta(\cos\theta - \cos\theta')\delta(\phi - \phi').
\end{equation}
This completes the proof of (\ref{toprove}).
Readers of Mukunda's paper \cite{mukunda1978completeness} should note that the proof therein of the completeness of the on-shell NRQM wavefunctions does not immediately carry over to the on-shell relativistic wavefunctions (\ref{eq:instate}, \ref{eq:outstate}). The last step of Mukunda's proof requires analysing the large-momentum behaviour of the wavefunctions in the complex momentum plane; this behaviour differs because in the non-relativistic case $\gamma=\alpha \frac{m}{|\vec{k}|}\rightarrow 0$, whereas the on-shell relativistic wavefunctions have an anomalous dimension that scales as $\gamma=\alpha\frac{E_p}{|\vec{p}|}\rightarrow \alpha$. We have found that, after taking this difference into account, one can modify the final steps of Mukunda's proof to establish completeness for the on-shell relativistic wavefunctions. However, we do not require this completeness relation in this paper.
\subsection{Cancellation of bound-state poles and retarded boundary conditions}\label{app:boundstatecancellation}
\subsubsection*{Placement of poles}
\begin{figure}
    \centering
    \includegraphics[scale=0.45]{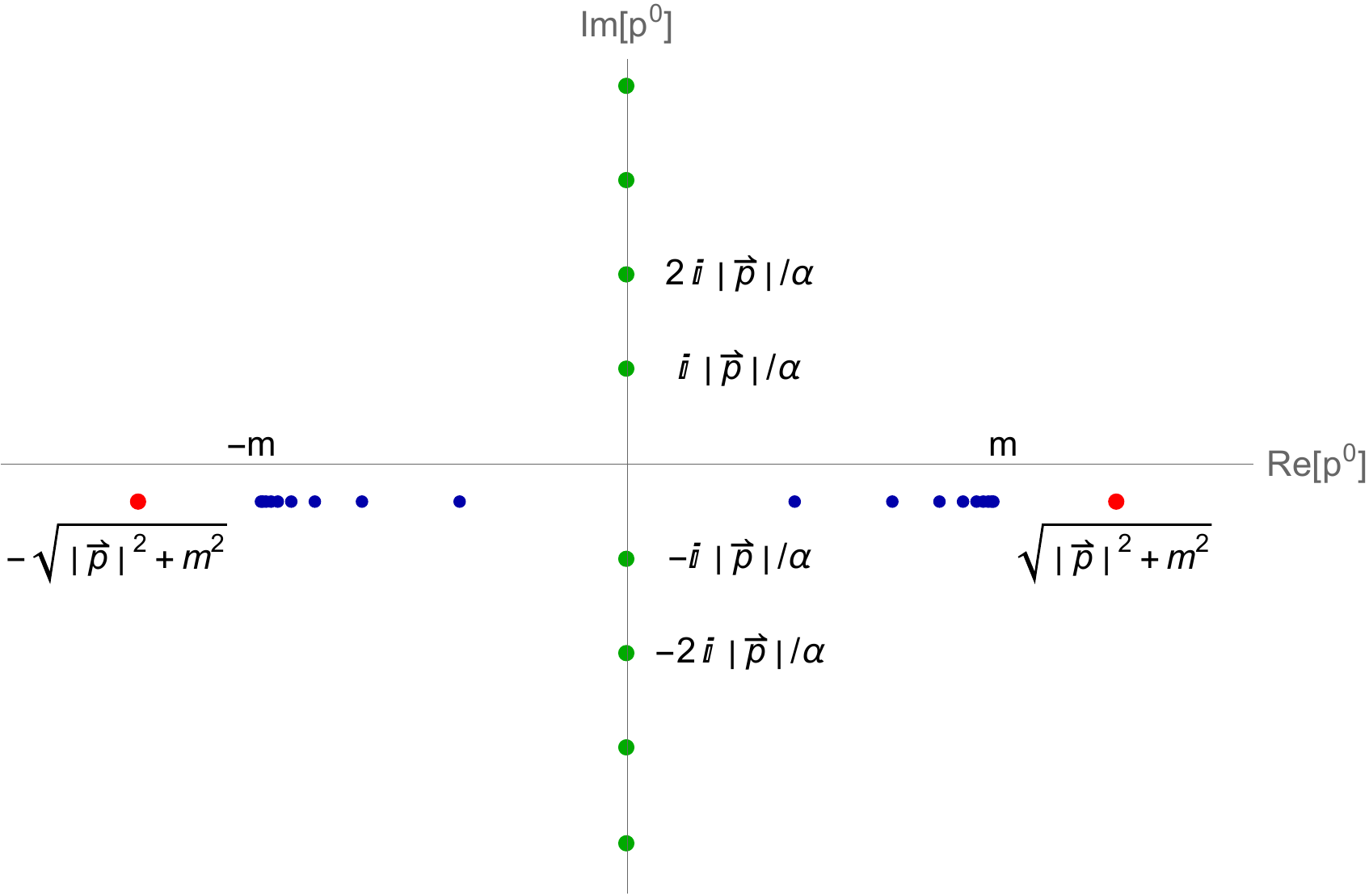}
    \caption{All of the poles of the integrand of the retarded Coulomb Green's function (\ref{greensforfig}) in the complex $p^0$ plane. The continuum wavefunctions $v_l$ contribute a factor of $\Gamma\left(1+l+i\alpha p^0/|\vec{p}|\right)\Gamma\left(1+l-i\alpha p^0/|\vec{p}|\right)$, which gives rise to simple poles along the imaginary axis at $p^0=\pm i n|\vec{p}|/\alpha$, $n=l+1,l+2,\ldots$, indicated in green. The denominator of the bound-state contribution to the Green's function (\ref{IBgreens}) gives rise to poles at $p^0=\pm m/\sqrt{1+\frac{\alpha^2}{n^2}}$, indicated in blue. We place these bound-state poles below the real axis to enforce retarded boundary conditions. The denominator of the continuum contribution to the Green's function contains two on-shell poles indicated in red, which we place below the real axis in order to impose retarded boundary conditions.}
    \label{fig:poles}
\end{figure}

We can decompose the retarded Green's function in terms of its spherical harmonic expansion as in (\ref{lotsasteps}),
\begin{align}
    G^{+}(x,y)&=\frac{1}{2\pi}\sum_{l=0}^{\infty}\sum_{m=-l}^lY_l^{m}(\hat{x})\Big( Y_l^{m}(\hat{y})\Big)^{\star}\Bigg(I_B+I^{+}_C\Bigg)\label{greensforfig}\\
    I_B&=\sum_{n=l+1}^{\infty}\int_{-\infty}^{\infty}\DD p^0 e^{-ip^0(x^0-y^0)} \theta(-\alpha p^0)\frac{\rho_{nl}\Big(r_x,\frac{1}{\alpha p^0}\Big)\rho^{\star}_{nl}\Big(r_y,\frac{1}{\alpha p^0}\Big)}{m^2-\Big(1+\frac{\alpha^2}{n^2}\Big)(p^0)^2}\label{IBgreens}\\
    I^{+}_C&=-\int_{-\infty}^{\infty}\DD p^0 e^{-ip^0(x^0-y^0)}\int_{0}^{\infty}\DD |\vec{p}| \,\, \frac{v_l\Big(|\vec{p}|,r_x,\frac{1}{\alpha p^0}\Big)v^{\star}_l\Big(|\vec{p}|,r_y,\frac{1}{\alpha p^0}\Big)}{p^2-m^2+i\e}\label{contgreen}
\end{align}
The continuum radial wavefunctions $v_l$ in (\ref{eq:contradial}) contribute a factor of $\Gamma(1+l+i\alpha \frac{p^0}{|\vec{p}|})\Gamma(1+l-i\alpha \frac{p^0}{|\vec{p}|})$ to the Green's function in (\ref{contgreen}), which gives rise to simple poles in the complex $p^0$ plane at
\begin{gather}
    p^0=\pm i n\frac{|\vec{p}|}{\alpha},\qquad n=l+1,l+2,\ldots\label{UHPpoles}
\end{gather}
The denominator of the bound state contribution to the Green's function (\ref{IBgreens}) gives rise to poles at
\begin{gather}
    p^0=\pm \frac{m}{\sqrt{1+\frac{\alpha^2}{n^2}}}.\label{boundstatepoles}
\end{gather}
We will show that in order to enforce retarded boundary conditions we must place these bound-state poles (\ref{boundstatepoles}) below the real $p^0$ axis. The final set consists of the usual on-shell $p^2=m^2$ poles in (\ref{contgreen}), which we place below the real $p^0$ axis to enforce retarded boundary conditions. These are all of the poles of the Green's function integrand. We plot these poles in Fig.~\ref{fig:poles}.
\subsubsection*{Cancellation of bound-state poles}
In order for the Green's function to satisfy retarded boundary conditions $G(x,y)\propto \theta(x^0-y^0)$, the bound-state poles in $I_C^{+}$ (\ref{UHPpoles}) in the upper half-plane (UHP) must be cancelled when including all contributions to the Green's function. In this section we show that a certain part of the bound-state contribution to the Green's function $I_B$ (\ref{IBgreens}) does indeed cancel these poles. \par
\begin{figure}
    \centering
    \includegraphics[scale=0.5]{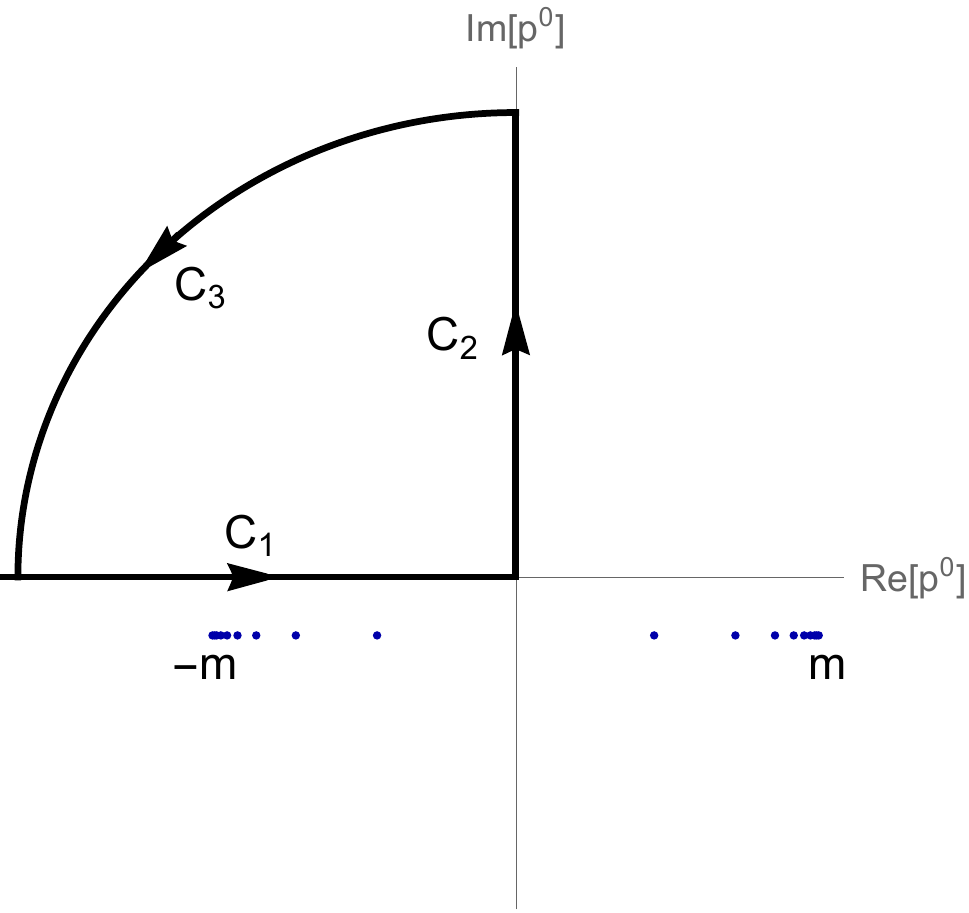}
    \caption{Contour deformation for $\alpha>0$ and $y^0>x^0$ for the $I_B$ integrand (\ref{IBgreens}), justifying the replacement in (\ref{bounddeform}). The original integration contour is $C_1$. Because the indicated quarter-circle contour encloses no poles, $C_1=-C_2-C_3$. Because $C_3=0$, we conclude that $C_1=-C_2$. For $\alpha<0$ the original integration contour $C_1'$ would instead be the half-line along the positive real axis $p^0>0$. Applying the same contour-deformation argument, we again obtain $C_2$, but with the opposite orientation to the previous one: $C_1'=C_2$.}
    \label{fig:halfarcdeform}
\end{figure}

Let us perform the $p^0$ integral in the continuum contribution to the Green's function $I_C^{+}$ (\ref{contgreen}). We consider the case $y^0>x^0$ first, in which case we close the $p^0$ integration contour in the UHP. The additional arc at infinity gives a zero contribution as the asymptotics are still dictated by the $e^{-ip^0(x^0-y^0)}$ factor. The integral is therefore given by the sum of the residues of the poles in the UHP. These poles arise from the Gamma functions in $v_l$ and have the following residue at the location of the pole
\begin{gather}
  \underset{p^0\rightarrow i\frac{|\vec{p}|}{|\alpha|}n }{\text{Res}}\Gamma\Bigg(1+l+i\alpha\frac{p^0}{|\vec{p}|}\Bigg)\Gamma\Bigg(1+l-i\alpha\frac{p^0}{|\vec{p}|}\Bigg)=i\frac{|\vec{p}|}{|\alpha|}(-1)^{n-l}\frac{(l+n)!}{(n-l-1)!}
\end{gather}
where we put absolute value brackets on $\alpha$ so we can simultaneously discuss $\alpha>0$ and $\alpha<0$. Picking up all of these poles in (\ref{contgreen}), recalling the definition of the $v_l$ functions (\ref{eq:contradial}), and performing some simplifications, we find
\begin{align*}
   &I_C(x,y)\Big\vert_{y^0>x^0}=(-1)^{l+1}\frac{4}{|\alpha|}\sum_{n=l+1}^{\infty}\frac{(l+n)!(4rr')^l}{(n-l-1)!\big[(2l+1)!\big]^2} \int_0^\infty\DD |\vec{p}| |\vec{p}|^{3+2l}  
e^{-i|\vec{p}|(|r_x|+|r_y|)} \\
&e^{n\frac{|\vec{p}|}{|\alpha|}(x^0-y^0)}\frac{{}_1F_1\Big(1+l+s(\alpha)n, 2l + 2; 2i|\vec{p}|r_x\Big){}_1F_1\Big(1+l+s(\alpha)n, 2l + 2; 2i|\vec{p}|r_y\Big)}{m^2+|\vec{p}|^2\Big(1+\frac{n^2}{\alpha^2}\Big)} \numberthis\label{eq:IC}
\end{align*}
where $s(\alpha)\defined \text{sign}(\alpha)$ arises in the first argument of the ${}_1F_1$ functions due to $-i\alpha \frac{p^0}{|\vec{p}|}\Big\vert_{p^0\rightarrow i\frac{|\vec{p}|}{|\alpha|}n}=s(\alpha)n$. Note that we used the relation
\begin{gather}
    {}_1F_1(a,b,-z)=e^{-z}{}_1F_1(b-a,b,z)\label{usefulconfidentity}
\end{gather}
to make the signs of the arguments of the confluent hypergeometric functions and the radial exponential factors the same.
We now demonstrate that (\ref{eq:IC}) is precisely cancelled by the $I_{B}$ contribution for $y^0>x^0$, thus ensuring retarded boundary conditions. The $\theta(-\alpha p^0)$ function in $I_B$ implies that $p^0$ runs only along half of the real line, and the sign of the half-line depends on the sign of $\alpha$. We will deform the contour as indicated in Fig.~\ref{fig:halfarcdeform}. The arc at infinity gives a zero contribution because $e^{-ip^0(x^0-y^0)}$ determines the asymptotics. Implementing the deformation described in Fig.~\ref{fig:halfarcdeform}, we find
\begin{gather}
    I_B=-\text{sign}(\alpha)\sum_{n=l+1}^{\infty}  \int_{0}^{i \infty}\DD p^0 e^{-ip^0(x^0-y^0)}\frac{\rho_{nl}\Big(r_x,\frac{1}{\alpha p^0}\Big)\rho^{\star}_{nl}\Big(r_y,\frac{1}{\alpha p^0}\Big)}{m^2-\Big(1+\frac{\alpha^2}{n^2}\Big)(p^0)^2}\label{bounddeform}
\end{gather}
where the integration contour is straight upwards along the imaginary $p^0$ axis. For each $n$ we change to a new integration variable from $p^0$ to $|\vec{p}|$, where
\begin{gather}
    p^0=\frac{in}{|\alpha|}|\vec{p}|
\end{gather}
This gives us
\begin{gather}
      I_B=-\text{sign}(\alpha)\sum_{n=l+1}^{\infty}\frac{in}{|\alpha|}  \int_{0}^{ \infty}\DD |\vec{p}| e^{\frac{n|\vec{p}|}{|\alpha|}(x^0-y^0)}\frac{\rho_{nl}\Big(r_x,in|\vec{p}|s(\alpha)\Big)\rho^{\star}_{nl}\Big(r_y,in|\vec{p}|s(\alpha)\Big)}{m^2+\Big(1+\frac{n^2}{\alpha^2}\Big)|\vec{p}|^2}
\end{gather}
where $s(\alpha)=\frac{\alpha}{|\alpha|}=\text{sign}(\alpha)$.
Plugging in the expression for $\rho$, (\ref{boundradial}) and simplifying we find
\begin{align*}
  &I_B= (-1)^l \frac{4}{|\alpha|}\frac{1}{ [(2l + 1)!]^2}  \sum_{n=l+1}^{\infty}\frac{(n + l)!}{(n - l - 1)!}\left( 4r r' \right)^l\int_{0}^{ \infty}\DD |\vec{p}|\,\, |\vec{p}|^{2l+3} e^{\frac{n|\vec{p}|}{|\alpha|}(x^0-y^0)}\\
  &e^{-is(\alpha)|\vec{p}|(r+r')}\frac{{}_1F_1\Big(l + 1 - n, 2l + 2, 2is(\alpha)|\vec{p}|r\Big){}_1F_1\Big(l + 1 - n, 2l + 2, 2is(\alpha)|\vec{p}|r'\Big)}{m^2+\Big(1+\frac{n^2}{\alpha^2}\Big)|\vec{p}|^2}\numberthis
\end{align*}
Upon using (\ref{usefulconfidentity}) to shift $s(\alpha)$ from the third argument of the ${}_1F_1$ functions to the first argument and the radial exponential factor, this expression precisely cancels (\ref{eq:IC}). We therefore have that the retarded Green's function with the placement of the poles shown in Fig.~\ref{fig:poles} does indeed satisfy retarded boundary conditions $G^{+}(x,y)\propto \theta(x^0-y^0)$.
\subsubsection*{The $x^0>y^0$ case}
To evaluate the $p^0$ integral in $I_{C}^{+}$ (\ref{contgreen}) for the case $x^0>y^0$, we can add an arc at infinity in the lower half-plane. This arc gives a zero contribution because the $e^{-ip^0(x^0-y^0)}$ factor determines the asymptotics. Therefore, $I_{C}^{+}$ for $x^0>y^0$ is given by the sum of its residues in the lower half-plane (LHP). These are the green ($p^0=-in\frac{|\vec{p}|}{|\alpha|}$) and red ($p^2=m^2$ on-shell) poles in Fig.~\ref{fig:poles}. For the half-line integral in the bound-state contribution (\ref{IBgreens}), we again perform a contour deformation, this time using a quarter-circle in the LHP analogous to Fig.~\ref{fig:halfarcdeform}. Doing so, we obtain a half-line contribution running along the imaginary $p^0$ axis. In addition, we also pick up the bound-state poles $(p^0)^2=m^2/(1+\frac{\alpha^2}{n^2})$ indicated in blue in Fig.~\ref{fig:poles}. As before, the green poles in $I_C^{+}$ are cancelled by the half-line running along the imaginary $p^0$ axis in $I_{B}$. The remaining contributions are the $p^2=m^2$ poles in $I_{C}^{+}$ and the bound-state poles (blue). We now calculate the residues at these two sets of poles. \par
The residues at these poles are as follows. First, consider the residue of the denominator term of the bound-state integrand (\ref{IBgreens}):
\begin{align}
    \underset{p^0\rightarrow \pm \frac{m}{\sqrt{1+\frac{\alpha^2}{n^2}}}}{\text{Res}}\frac{1}{m^2-\Big(1+\frac{\alpha^2}{n^2}\Big)(p^0)^2}&=\mp \frac{1}{2m}\frac{1}{\sqrt{1+\frac{\alpha^2}{n^2}}}\\
    &=\text{sign}(\alpha) \frac{|E_n|}{2m^2}
\end{align}
The pole we pick up is dictated by the sign of $\alpha$. If $\alpha$ is positive then we pick up the negative pole, and vice versa. The orientation is clockwise in both cases, thus giving a factor of $-(2\pi i)$. Altogether the bound-state contribution is given purely in terms of its on-shell expansion
\begin{gather}
  G^{+}_B(x,y)=i\,   \theta(x^0-y^0)\sum_{l=0}^{\infty}\sum_{n=l+1}^{\infty}\sum_{m=-l}^lE_n\phi_{nlm}(x)\phi^{\star}_{nlm}(y)
\end{gather}
Here we have returned to the form (\ref{greensforfig}) and restored the spherical-harmonic factors. We used the factor $\sqrt{2}m$ relating the NRQM and relativistic wavefunctions in (\ref{boundstatesintermsofNRQM}), together with the fact that the on-shell functions reduce to $\phi_{nlm}$. Next we consider the continuum contribution $I_{C}^{+}$. Since we have demonstrated that the gamma-function poles cancel, we can return to the form (\ref{continiuumgreens})
\begin{gather}
    G^{+}_C(x,y)=-\int\frac{\DD^4p}{(2\pi)^4} \,\, \frac{\chi_{\text{in/out}}(x,p,u_s)\chi_{\text{in/out}}^{\star}(y,p,u_s)}{p^2-m^2+i\e}
\end{gather}
and pick up only the $p^2=m^2$ on-shell poles, to obtain
\begin{gather}
     G^{++}_C(x,y)=i\theta(x^0-y^0)\int\frac{\DD^3 p}{(2\pi)^3}\frac{1}{2E_p} \Bigg(\phi_{\text{in/out}}(x,p,u_s)\phi_{\text{in/out}}^{\star}(y,p,u_s)-(E_p\rightarrow -E_p)\Bigg)
\end{gather}
where we have added an additional plus sign to the superscript as this is not strictly equal to $G^{+}$, due to the required addition of the off-shell bound-state Green's function to enact the cancellation of the Gamma-function poles.
Altogether we have
\begin{align*}
    G^{+}(x,y)=i\, &\theta(x^0-y^0)\Bigg\lbrace
    \sum_{l=0}^{\infty}\sum_{n=l+1}^{\infty}\sum_{m=-l}^lE_n\phi_{nlm}(x)\phi^{\star}_{nlm}(y)\\
    &+\int\frac{\DD^3 p}{(2\pi)^3}\frac{1}{2E_p} \Big(\phi_{\text{in/out}}(x,p,u_s)\phi_{\text{in/out}}^{\star}(y,p,u_s)-(E_p\rightarrow -E_p)\Big)\numberthis
    \Bigg\rbrace
\end{align*}
For the case of the advanced Green's function, the argument of the theta function changes sign, and we obtain an overall minus sign due to the change in orientation when picking up the on-shell residues.
\subsection{Vanishing of the equal-time retarded Green's function}\label{app:vanishretardedgreens}
Following Eqs. (2.12)--(2.16) of Hostler \cite{1964JMP.....5..591H}, we will show that the retarded Green's function vanishes at equal times
\begin{align*}
\delta(x^0-y^0) \Bigg\lbrace   &\sum_{l=0}^{\infty}\sum_{n=l+1}^{\infty}\sum_{m=-l}^lE_n\phi_{nlm}(x)\phi^{\star}_{nlm}(y)\\
&+\int\frac{\DD^3p}{(2\pi)^32E_p}\Bigg(\phi_{\text{in/out}}(p,x)\phi^{\star}_{\text{in/out}}(p,y)-(E_{p}\rightarrow -E_p)\Bigg)\Bigg\rbrace=0\label{toproves}\numberthis
\end{align*}
Hostler considered the Green's function in the partial-wave basis, and furthermore considered the solutions of the full SQED equations of motion. For ease of reference, and to adapt the proof to our conventions for the normalization of the states, we reproduce their proof with minor modifications here.\par
Assuming, without proof, that the continuum and bound-state solutions form a complete basis for the solution space of the semi-free EQM, we have that any SQED-normalizable solution of the semi-free EQM can be expressed in this basis
\begin{align*}
    f(x)=&\sum_{l=0}^{\infty}\sum_{n=l+1}^{\infty}\sum_{m=-l}^lb_{nlm}E_n\phi_{nlm}(x)\\
    &+\int\frac{\DD^3 k}{(2\pi)^32E_k}\,\,  \Bigg(a_{+}(k)\phi_{\text{in/out}}(k,x)-a_{-}(k)\phi_{\text{in/out}}(\vec{k},-E_k,x)\Bigg),\label{measure}
    \numberthis
\end{align*}
where the coefficients can be extracted using the orthogonality (\ref{orthonormal}, \ref{boundortho}) of the basis functions:
\begin{gather}
    b(n,l,m)=\braket{\phi_{nlm}|f}_{\text{SQED}}\qquad a_{\pm}(k)=\big\langle\phi_{\text{in/out}}(k,\pm E_k)\big\vert f\big\rangle _{\text{SQED}}.
\end{gather}
The measure in (\ref{measure}) is chosen so as to compensate for the norm of the states.  Let us consider the expansion (\ref{measure}) at time $t=0$, and use the explicit expressions for the expansion coefficients
\begin{align*}
&f(\vec{x},t=0)=\Bigg[\sum_{l=0}^{\infty}\sum_{n=l+1}^{\infty}\sum_{m=-l}^lE_n\int\DD^3 y\phi_{nlm}(x)\phi_{nlm}^{\star}(y)\Bigg(i\dot{f}(\vec{y},0)+\Big(iE_n-\frac{\alpha}{r}\Big)f(\vec{y},0)\Bigg)\\
    &+\int\frac{\DD^3 y\,\DD^3 k}{(2\pi)^32E_k}\phi_{\text{in/out}}(k,x)\phi^{\star}_{\text{in/out}}(k,y)  \Bigg(i\dot{f}(\vec{y},0)+\Big(iE_k-\frac{\alpha}{r}\Big)f(\vec{y},0)\Bigg)\\
    &-\int\frac{\DD^3 y\,\DD^3k}{(2\pi)^32E_k}\phi^{-E_k}_{\text{in/out}}(k,x)\phi^{\star,-E_k}_{\text{in/out}}(k,y)  \Bigg(i\dot{f}(\vec{y},0)+\Big(-i|E_k|-\frac{\alpha}{r}\Big)f(\vec{y},0)\Bigg)\Bigg]\Bigg\vert_{x^0=y^0=0}.\numberthis\label{Hostleridentity}
\end{align*}
Now, because the SQED EQM are second-order in time, $f(\vec{x},0)$ can be prescribed arbitrarily, independently of $\dot{f}(\vec{x},0)$. Thus $\dot{f}$ must drop out of (\ref{Hostleridentity}). To make this argument precise, choose $f(\vec{x},0)=0$ in (\ref{Hostleridentity}), which gives us an entire function space of identities relating the $\phi$ functions, one for each choice of $\dot{f}(\vec{y},0)$. These relations can only be satisfied if the coefficient of $\dot{f}(\vec{y},0)$ in (\ref{Hostleridentity}) vanishes at each point in $y$-space, namely
\begin{align*}
\delta(x^0-y^0)&\Bigg[\sum_{l=0}^{\infty}\sum_{n=l+1}^{\infty}\sum_{m=-l}^lE_n\phi_{nlm}(x)\phi_{nlm}^{\star}(y)\\
    &+\int\frac{\DD^3 k}{(2\pi)^32E_k}\Bigg(\phi_{\text{in/out}}(k,x)\phi^{\star}_{\text{in/out}}(k,y)-\phi^{-E_k}_{\text{in/out}}(k,x)\phi^{\star,-E_k}_{\text{in/out}}(k,y)\Bigg)\Bigg]=0,\label{proved}\numberthis
\end{align*}
which is what we sought to prove (\ref{toproves}). Now that we have established that the $\dot{f}(\vec{y},0)$ term drops out of (\ref{Hostleridentity}), we can also derive a completeness relation. Note that the $\frac{\alpha}{r}$ terms in (\ref{Hostleridentity}) also vanish on account of (\ref{proved}). Now, as $f(\vec{x},0)$ occurs on the LHS and RHS of (\ref{Hostleridentity}), the integral over $\vec{k}$-space of the coefficients of $f(\vec{y},0)$ in the remaining terms on the RHS must equal $\delta^3(\vec{x}-\vec{y})$,
\begin{align*}
&i\, \delta(x^0-y^0)\Bigg[\sum_{l=0}^{\infty}\sum_{n=l+1}^{\infty}\sum_{m=-l}^lE^2_n\phi_{nlm}(x)\phi_{nlm}^{\star}(y)\\
    &+\frac{1}{2}\int\frac{\DD^3 k}{(2\pi)^3}\Bigg(\phi_{\text{in/out}}(k,x)\phi^{\star}_{\text{in/out}}(k,y)  +\phi^{-E_k}_{\text{in/out}}(k,x)\phi^{\star,-E_k}_{\text{in/out}}(k,y)\Bigg)\Bigg]=\delta^4(x-y)\numberthis\label{completes}
\end{align*}
which indicates the completeness of the continuum and bound states.
The argument that led to (\ref{completes}) is circular, as we assumed without proof the completeness of the bound-state and continuum solutions in (\ref{measure}), and then consistently concluded that the states form a complete set. We consider the completeness assumption to be reasonable, and it is nevertheless useful to present the expected form of the completeness relation (\ref{completes}) for future work that may prove the relation by other means.
\bibliography{coulomb.bib}
\bibliographystyle{JHEP}
\end{document}